\documentclass[10pt,twocolumn, pra, aps, longbibliography, superscriptaddress]{revtex4-2}
\usepackage{amssymb, amsmath, amsfonts}
\usepackage[utf8]{inputenc}   
\usepackage[T1]{fontenc}
\usepackage[]{inputenc}
\usepackage{color}
\usepackage{amstext}
\usepackage{enumitem}
\usepackage{comment}
\usepackage{graphicx, bm, palatino}

\usepackage{microtype}
\usepackage[dvipsnames]{xcolor}
\usepackage{xparse}

\usepackage[sc]{mathpazo}
\usepackage{tikz}
\usetikzlibrary{shapes.geometric, arrows.meta, positioning, decorations.pathmorphing,shapes}
\usepackage{placeins}

\DeclareMathOperator{\Tr}{Tr}
\usepackage{times}
\usepackage{bbm}

\usepackage[colorlinks=true, linkcolor=blue, urlcolor=blue, citecolor=blue, pdfusetitle]{hyperref}
\usepackage{cleveref}
\begin{document}
\title{Non-equilibrium quantum thermodynamics of a memory-bearing open-system process}

\author{Biagio G. Banigi}
\affiliation{Universit\`a degli Studi di Palermo, Dipartimento di Fisica e Chimica - Emilio Segr\`e, via Archirafi 36, I-90123 Palermo, Italy}
\affiliation{Departamento de Física Teórica, Facultad de Ciencias Físicas, Universidad Complutense, 28040 Madrid, Spain}

\author{Eric Lutz}
\affiliation{Institute for Theoretical Physics I, University of Stuttgart, D-70550 Stuttgart, Germany}

\author{Mauro Paternostro}
\affiliation{Universit\`a degli Studi di Palermo, Dipartimento di Fisica e Chimica - Emilio Segr\`e, via Archirafi 36, I-90123 Palermo, Italy}
\affiliation{Centre for Quantum Materials and Technologies, School of Mathematics and Physics, Queen's University Belfast, BT7 1NN, United Kingdom}

\begin{abstract}
{We show the emergence of memory effects in the dynamics of a driven two-level system interacting with a composite environment, and analyze their influence on work, heat and entropy production. We further investigate how the interplay between driving, dissipation and memory effects, stemming from the finiteness of the environment, shapes the thermodynamic response of the system, thus providing  insight into quantum thermodynamics beyond the Markovian approximation.}
\end{abstract}

\maketitle

\section{INTRODUCTION}\label{section:introduction}
{
The description of the open system dynamics of a quantum system resulting from the coupling with  its surroundings has often relied on the Markovian approximation, which forces information to monotonically flow from the former to the latter ~\cite{RevModPhys.88.021002}. Under such assumption, the evolution of the state of the system can be depicted by means of positive quantum dynamical maps ~\cite{Lindblad,GKS}. 
However, there might be situations - such as in the presence of finite or structured environments- in which a sharp separation between timescales of system and environment is missing. This makes a Markovian treatment unsuitable to describe properly the evolution of a system~\cite{Breuer,PhysRevResearch.2.043006}. In some cases, an information backflow from the environment to the system may occur, and memory effects remarkably alter the system dynamics, entering a non-Markovian regime ~\cite{PhysRevLett.103.210401, Rivas_2014, RevModPhys.89.015001}}.

{A potentially fruitful step to make in order to assess this particular class of open quantum system dynamics is to reconsider it from the perspective of thermodynamics. In this sense, the redefinition of quantities as heat, work and energy remains crucial for the construction of thermodynamics laws for quantum systems~\cite{Binder2018}. Similarly, a deep analysis of entropy production, and its rate of change, represents the key for the understanding of irreversible processes at the microscopic level ~\cite{PhysRevResearch.2.043006}. A significant body of work has been produced  in the memoryless case, where many results from classical thermodynamics could be retrieved~\cite{Spohn, PhysRevResearch.2.043006}.}

{More recently, the thermodynamic implications of quantum systems undergoing a non-Markovian evolution have been actively studied, particularly in regard to the presence of negative entropy production rates~\cite{Marcantoni,Steve,PhysRevE.99.012120}. Despite these advances, the combined effects of non-Markovianity with driving and  dissipation in non-equilibrium contexts~\cite{Dann2024, follia2025controlmemoryeffectsspinboson,PhysRevResearch.6.013258,Abah_2020} still embody an active area of investigation.}

{Motivated by the need to understand how these phenomena coexist and influence both the dynamics and thermodynamics of open quantum systems,  we study a non-Markovian setup, in which a two-level system interacts with a composite environment. We start by considering a harmonic oscillator as a finite-size environment, and further extend the discussion by including a bosonic bath and an external driving}. 
This allows us to analyze the interplay among non-equilibrium phenomena and the role they play in the dynamics and thermodynamics of the system.
To this purpose, we study the evolution of the density matrix of the qubit, by developing a Lindblad-like master equation. 

We then proceed with a thermodynamic analysis within the formalism of open systems, consisting in the investigation of the heat exchanged by the qubit with its environment, the performed work and the entropy production. The latter, representing the fundamental tool for the investigation of non-equilibrium properties, can be used to work out aspects of the interchange between driving, dissipation, and memory effects and how it is possible to enhance or hinder their reciprocal interactions by appropriately tuning the coupling strengths \cite{PhysRevLett.107.140404, doi:https://doi.org/10.1002/9780470142578.ch2, lindblad1983nonequilibrium}.

{The remainder of this paper is organized as follows. In Sec.~\ref{section:physicalmodel} we introduce the physical setup and start its analysis. In detail, in Sec.~\ref{section:nonMarkovianity} we discuss the measure used to investigate the presence of non-Markovianity in the system evolution, while the general thermodynamic framework is outlined in Sec.~\ref{section:nonMarkovianthermo}. Then, in Sec.~\ref{section:results} dynamics and thermodynamics results are presented (Secs.~\ref{section:results_dynamics} and~\ref{section:results_thermodynamics}, respectively). Finally, our conclusions are summarized in Section~\ref{section:conclusions}.}

\begin{figure}[b!]
			\centering
			 \def\myscale{0.83}
    
                \begin{tikzpicture}[
                    thick,
                    >=Stealth,
                    scale=\myscale,
                    transform shape
                ]
				
				\node[circle, draw, minimum size={3cm*\myscale}, 
				fill=red!20, label=below:Two-level system] (system) at (0,0) {};

				\draw[very thick] (-0.7*\myscale, 0.4*\myscale) -- (0.7*\myscale, 0.4*\myscale) node[right] {$|e\rangle$};
				\draw[very thick] (-0.7*\myscale, -0.4*\myscale) -- (0.7*\myscale, -0.4*\myscale) node[right] {$|g\rangle$};

                \draw[-{Stealth[length=2mm*\myscale,width=1.5mm*\myscale]}, black, thick] (0.08*\myscale, -0.30*\myscale) -- (0.08*\myscale, 0.30*\myscale) node[midway, right] {$\omega_S$};
		      \draw[-{Stealth[length=2mm*\myscale,width=1.5mm*\myscale]}, black, thick] (0.08*\myscale, 0.30*\myscale) -- (0.08*\myscale, -0.30*\myscale);

				\node[label=below:Harmonic oscillator] (oscillator) at (-4.5*\myscale, -1.5*\myscale) {};

				\draw[thick, green!70!black] (-6.2*\myscale, 1.8*\myscale) parabola bend (-4.5*\myscale, -1.5*\myscale) (-2.8*\myscale, 1.8*\myscale);

				\foreach \n in {0, 1, 2, 3} {
					\pgfmathsetmacro{\ypos}{(-0.70+0.60*\n)*\myscale}
					\pgfmathsetmacro{\xwidth}{(0.45*sqrt(2+0.8*\n))*\myscale}

					\draw[thick, black] ({-4.5*\myscale-\xwidth}, \ypos) -- ({-4.5*\myscale+\xwidth}, \ypos);
				}

                \draw[-{Stealth[length=2mm*\myscale,width=1.5mm*\myscale]}, black, thick] (-4.5*\myscale, -0.1*\myscale) -- (-4.5*\myscale, 0.5*\myscale) node[midway, right] {$\omega_{ho}$};
                
                \draw[-{Stealth[length=2mm*\myscale,width=1.5mm*\myscale]}, black, thick] (-4.5*\myscale, 0.5*\myscale) -- (-4.5*\myscale, -0.1*\myscale);

				\node at (-4.5*\myscale, 1.6*\myscale) {$\vdots$};

				\draw[<->,  green!70!black , very thick] (-3.2*\myscale, 0) -- (-1.5*\myscale, 0) 
				node[midway, above] {$J$};

				\node[rectangle, draw, minimum width=3cm *\myscale, minimum height=4cm *\myscale, 
				fill=blue!20, label=below:Thermal bath] (bath) at (4.5*\myscale, 0) {};

				\foreach \i in {1,2,3} {
					\draw[thick, red, decoration={snake, amplitude=1mm}, decorate] 
					(3.7*\myscale +0.2*\i *\myscale, -0.3 *\myscale +0.2*\i *\myscale) -- (5.3*\myscale -0.5*\i*\myscale, 0.3 *\myscale -0.2*\i *\myscale);
				}

				\node[rectangle, draw, minimum width=3cm*\myscale, minimum height=4cm*\myscale,  
				fill=blue!20, label=below:Thermal bath] (bath) at (4.5*\myscale, 0) {};

				\draw[<->, blue, very thick] (1.5*\myscale, 0) -- (3.5*\myscale, 0) 
				node[midway, above] {$\Gamma$};

				\node[label=above:External drive, text=orange] (drive) at (1.4*\myscale, 2.0*\myscale) {$\lambda$};

				\draw[->, orange, very thick, line width=1pt*\myscale, 
				decorate, decoration={snake, amplitude=0.8mm*\myscale, segment length=2.7mm*\myscale, post length=2mm*\myscale}] 
				(-0.7*\myscale, 3.5*\myscale) -- (0, 1.6*\myscale);
			\end{tikzpicture}
			\caption{Sketch of the open system process described in Sec.~\ref{section:physicalmodel}. A two-level system is coupled to a harmonic oscillator (with strength $J$) and a thermal bath ($\Gamma$), and it is driven by an external field (described by the parameter $\lambda$).}
			\label{fig:two_level_System}
		\end{figure}
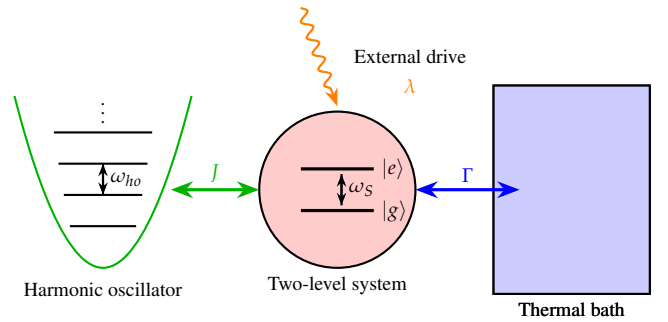

\section{PHYSICAL MODEL and tools}\label{section:physicalmodel}

Let us consider a single two-level system with characteristic frequency $\omega_S$. Adopting the natural units $\hbar=1$, its Hamiltonian can be written as
\begin{equation}
    H_S = \frac{ \omega_S}{2} \sigma_z,
\end{equation}
where $\sigma_z$ is the $Z$-Pauli matrix.
The qubit interacts with a harmonic oscillator at a frequency $\omega_{ho}$ with Hamiltonian
\begin{equation}
    H_{ho} = \omega_{ho} \bigg(a^{\dagger}a + \frac{1}{2}\bigg),
\end{equation}
where $a$ and $a^{\dagger}$ are the ladder operators. In this setup, the qubit-oscillator coupling will be given by the Jaynes-Cummings Hamiltonian under the rotating-wave approximation~\cite{raimond2006exploring}
\begin{equation}\label{equation:linearIqubit}
    H_I = J (a\sigma_+ + a^{\dagger}\sigma_-).
\end{equation}
We assume the possibility of switching on an external field which influences the qubit's dynamics. This means we need to add a Hamiltonian term which takes into account the driving protocol.
For a qubit, the most general form we can provide for such a Hamiltonian is $H_D (t) =  f(t) \ \sigma_x + g(t)\ \sigma_y  + h(t) \ \sigma_z$, with $f(t)$, $g(t)$ and $h(t)$ being time-dependent functions. 
{A particularly interesting case is that of a two-level system driven by a rotating magnetic field, as investigated in Ref.~\cite{Denzler_2024}. Motivated by this scenario, particularly interesting for its experimental value ~\cite{PhysRevLett.123.240601}, we take} $f(t) = \lambda \cos(\omega_D t)$, $g(t) = \lambda \sin(\omega_D t)$ and $h(t)=0$, {so that  $H_D(t)$ reads}
\begin{equation}
	 H_D (t)=  \lambda [\cos(\omega_D t)\sigma_x +  \sin(\omega_D t)\sigma_y].
\end{equation}
In this form, $H_D (t)$ captures the same physical mechanism considered in Ref.~\cite{Denzler_2024}. 
The generator of the dynamics thus reads 
\begin{equation}
    H(t) = H_S + H_D(t) + H_{ho} + H_I.
\end{equation}
{In order to model the reduced evolution of the qubit, we start  assuming that the state of the $S+E$ compound is initially factorized as $\rho_{SE} (0)=\rho_{S}(0) \otimes \rho_{E}(0)$. Here, $\rho_{S}(0)$ is a generic state of the two-level system and $\rho_{E}(0) = |p \rangle \langle p|$ is a state of the oscillator with exactly $p$ excitations. Following the approach developed in Ref.~\cite{Andersson15082007}, we find a Lindblad-like master equation for the dynamics of the qubit (the full derivation is reported in Appendix~\ref{appendix1}). In the interaction picture, the equation reads $\dot{\rho}_S(t)= -i[H_D(t), \rho_S (t)] +\mathcal{D}_t[{\rho_S(t)}]$, with the dissipator 
\begin{equation}
\label{eq:ME}
\mathcal{D}_t[{\rho_S(t)}]=\sum^3_{j=1}\gamma_j(t)\left(L_j\rho_S(t)L^\dag_j-\frac12\left\{L^\dag_jL_j,\rho_S(t)\right\}\right),
\end{equation}
where we have introduced the jump operators $L_1=L_2^\dag=i\sigma_-$, $L_3=\sigma_z$ and the (time-dependent) rates $\gamma_j(t)$.}

We also consider the effects of a memoryless bosonic bath, directly interacting with the qubit. The corresponding physical configuration is shown in Fig.~\ref{fig:two_level_System}.
{The inclusion of the reservoir is performed within the weak-coupling regime, where the strength $\Gamma$ of the qubit-bath interaction is sufficiently small compared to the qubit–oscillator coupling $J$. Under this assumption, it is possible to treat the degrees of freedom of the harmonic oscillator separately from those of the thermal bath. This allows one to incorporate the effects of dissipation directly at the level of the master equation by means of two additional terms,
accounting for the exchange of excitations between the reservoir and the qubit.
These contributions have the same operational part as the first two in Eq.~\eqref{eq:ME}, but with time-independent rates, related to $\Gamma$ and the average number of excitations in the reservoir $n$, found from the Planck distribution~\cite{Breuer}. Accordingly, we can rewrite Eq.~\eqref{eq:ME} by replacing $\gamma_1(t) \to \gamma_1(t) + \Gamma (n+1)$ and $\gamma_2(t) \to \gamma_2(t) + \Gamma n$.
However, care must be taken when choosing the simulation parameters to ensure the physical consistency of the dynamics. In particular, when the qubit–bath coupling strength $\Gamma$ becomes comparable to $J$, this \textit{phenomenological} description may lead to unphysical outcomes. For this reason, the results discussed throughout this work are obtained in the weak-coupling regime, with $\Gamma$ taken at least one order of magnitude smaller than $J$.}

\subsection{Probing non-Markovianity}\label{section:nonMarkovianity}
The tool that we use to characterize the dynamics of our system 
pivots around the  
so-called BLP (Breuer-Laine-Piilo) measure~\cite{PhysRevLett.103.210401}. 
Given two states $\rho_{1,2}$ of the system, their norm-1 distance reads 
\begin{equation}
	D(\rho_1,\rho_2) = \frac{1}{2} \Tr|\rho_1-\rho_2|\in[0,1]
    \label{eq:tracedistance}
\end{equation}
\noindent with $|O|=\sqrt{O^{\dagger}O}$. We have $D(\rho_1,\rho_2) = 0$  ($D(\rho_1,\rho_2) = 1$) if $\rho_1=\rho_2$ (if the states are orthogonal).
The trace distance is {contractive} under any completely positive (CP) memoryless process $\phi_t = \phi(t,0)$~\cite{1994RvMaP...6.1147R}, that is
\begin{equation}
	D \bigg(\phi_t \big[ \rho_1 \big],\phi_t \big[\rho_2 \big] \bigg) \le D(\rho_1,\rho_2) .
	\label{eq:contractivedistance}
\end{equation}
By using the semigroup property of divisibility $\phi_{t+\tau}=\phi_{\tau}\phi_t$ ~\cite{RevModPhys.88.021002}, with $\phi_{\tau} = \phi(\tau + t,t)$ and $\phi_{\tau+ t} = \phi(\tau + t,0)$, we can write Eq.~\eqref{eq:contractivedistance} for any $t, \tau \ge 0$ as
\begin{equation}
	D \bigg(\rho_1( \tau + t),\rho_2(\tau  + t) \bigg) \le D \bigg(\rho_1(t),\rho_2(t) \bigg) ,
	\label{eq:timedepdivisibility}
\end{equation}
where $\rho_j(t) = \phi_t[\rho_j(0)]$. 
We consider the rate of the trace distance
\begin{equation}
	\sigma(t,\rho_{1,2}(0))=\frac{d}{dt}D \bigg(\rho_1(t),\rho_2(t) \bigg), 
    \label{eq:tracedistancerate}
\end{equation}
which is overall negative for the reasons seen so far. As remarked in Ref.~\cite{PhysRevLett.103.210401}, Eq.~\eqref{eq:contractivedistance} holds not only for CP maps but for the larger class of positive and trace-preserving maps. Thus, $\sigma(t,\rho_{1,2}(0)) \le 0$ strictly requires the intermediate map $\phi_{\tau}$ to be just positive (P-divisibility). These relations identify a dynamics where a progressive loss of distinguishability between the initial states occurs.
However, we also give to this property a different interpretation: the loss of distinguishability is a consequence of the coupling between system and environment. This bond creates an information flow from the first to the latter, which hinders the possibility of distinguishing  $\rho_1 $ and $\rho_2$~\cite{PhysRevLett.103.210401}.
Such a phenomenology holds true for time-dependent Lindblad-like master equations where Hamiltonian, rates and jump operators might all depend on time. For non-negative time-dependent rates, the corresponding map remains CP-divisible, and consequently Eq.~\eqref{eq:timedepdivisibility} still holds
~\cite{PhysRevLett.103.210401, RevModPhys.88.021002}.

When the non-negativity of the rates is broken, P-divisibility is not guaranteed anymore and  $D(\rho_1,\rho_2)$ could exhibit revivals, or in general a non-monotonic behavior. Correspondingly, also the flow of information reverses, and we witness a \textit{backflow} of information from the environment to the system~\cite{PhysRevLett.103.210401}. This is the case of Eq.~\eqref{eq:ME} we obtained previously.

\subsection{Thermodynamics beyond the Markovian regime}\label{section:nonMarkovianthermo}
We now introduce the thermodynamic framework with the definition of work and heat. In the weak-coupling regime, which we assume throughout, the standard expressions are given in terms of rates ~\cite{Alicki_1979, Gemmer2009, Rivas2019}

\begin{gather}
	\frac{dW(t) }{dt} = \Tr \bigg\{\rho_S(t) \frac{dH(t) }{dt} \bigg\}, \label{eq:derwork} \\
    \frac{dQ(t) }{dt} = \Tr \bigg\{\frac{d\rho_S(t) }{dt} H(t) \bigg\} .\label{eq:derheat}
\end{gather}

{The latter, in particular, naturally enters the formulation of the second law through the Clausius inequality
\begin{equation}
	dS(t)\ge \beta dQ(t),
	\label{eq:clausiusinequality}
\end{equation}
with $\beta$ the inverse temperature of the environment and $S$  the thermodynamic entropy, which in the quantum regime is usually identified as the von Neumann entropy $S= -\Tr \{ \rho_S \ln (\rho_S) \}$.
 The equality holds for reversible processes, when the only contribution to entropy comes from the exchange of heat between the system and the surroundings. In general, one can introduce the entropy production ~\cite{Marcantoni, Gemmer2009, Rivas2019, Esposito_2010}
\begin{equation}
	\Sigma (t)= \Delta S(t) - \beta Q(t).
    \label{eq:entropyproddefinition}
\end{equation}

The second law enforces $\Sigma(t) \ge 0$. Eq.~\eqref{eq:entropyproddefinition} can be  rewritten as the rate equation 
\begin{equation}
    \sigma_{\Sigma}(t) \equiv \frac{d\Sigma(t)}{dt} = \frac{dS(t)}{dt} -  \Phi(t),
    \label{eq:entropybalance}
\end{equation}
where $\Phi(t)=d(\beta Q(t))/dt$ identifies the entropy flux from the system to the environment ~\cite{PhysRevResearch.2.043006,Steve2}. Thus, for a system evolving towards a non-equilibrium steady state $\bar{\rho}_s$, associated with the long-time dynamics generated by $\mathcal{L}$, the entropy production rate reads ~\cite{Spohn}
\begin{equation}
	\sigma_{\Sigma}(t) = \Tr \{ \mathcal{L}(\rho_s (t)) [ \ln(\bar{\rho}_s) - \ln(\rho_s(t)) ] \}.
\end{equation}
The non-negativity of $\sigma_{\Sigma}(t)$ is guaranteed from the so-called Spohn's inequality, which holds whenever we deal with P-divisible dynamics~\cite{Steve2,Steve}.

{However, as discussed in Sec.~\ref{section:nonMarkovianity}, not every physical process requires divisibility. In particular, the lack of P-divisibility of the dynamical map can be connected to the emergence of non-Markovianity and the development of distinguishability between states of the system.} {As a consequence, in systems evolving according to a non-Markovian dynamics,  entropy production may decrease, and even become transiently negative.  
Such behavior may appear as a counterintuitive violation of the second law of thermodynamics. A more careful analysis requires a reformulation of the second law that addresses explicitly the non-negligible role of the environment. The underlying reason is connected to its dissipative dynamics and the emergence of the information backflow that non-Markovianity entails~\cite{Marcantoni, PhysRevE.99.012120}.}

{In the following, we exploit this framework to analyze how non-Markovian dynamics influences the thermodynamics properties of our composite system.}

\section{RESULTS}\label{section:results}
In this Section, we present and discuss some numerical results following from the {introduction of the model and the description of tools performed} 
in Sec.~\ref{section:physicalmodel}. We proceed by separating dynamical [cf. Sec.~\ref{section:results_dynamics}] and thermodynamical [Sec.~\ref{section:results_thermodynamics}] implications of the physical configuration at hand.

\subsection{Non-Markovian Dynamics}\label{section:results_dynamics}
We now analyze the dynamical properties of the qubit  arising from the interaction with the structured environment and the external driving. 
Particular attention is devoted to the emergence of memory effects, whose impact can be described via Eq.~\eqref{eq:tracedistance} and Eq.~\eqref{eq:tracedistancerate}. To characterize the temporal behavior of the trace distance 
{$\rho_1(0) = | + \rangle \langle +|$ and $\rho_2(0) = | - \rangle \langle - |$, where $| \pm\rangle = (| e \rangle \pm |g\rangle)/\sqrt{2}$ and $| e \rangle$ and $|g \rangle$ are, respectively, the excited and ground state.} 
\begin{figure}[t!]
\centering
\makebox[\columnwidth][l]{\hspace{0.01\columnwidth}\textbf{(a)}}\\
\includegraphics[width=\columnwidth]{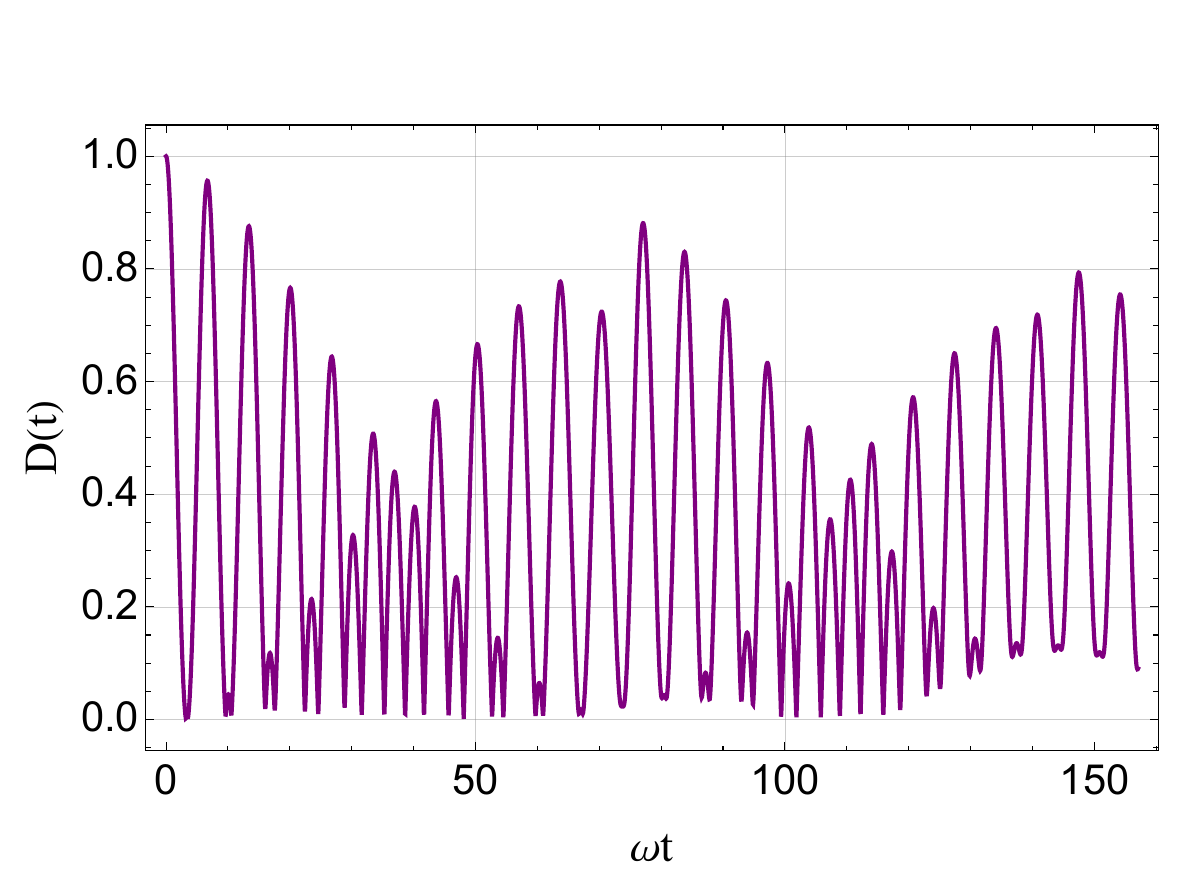}\\

\makebox[\columnwidth][l]{\hspace{0.01\columnwidth}\textbf{(b)}}\\
 \includegraphics[width=\columnwidth]{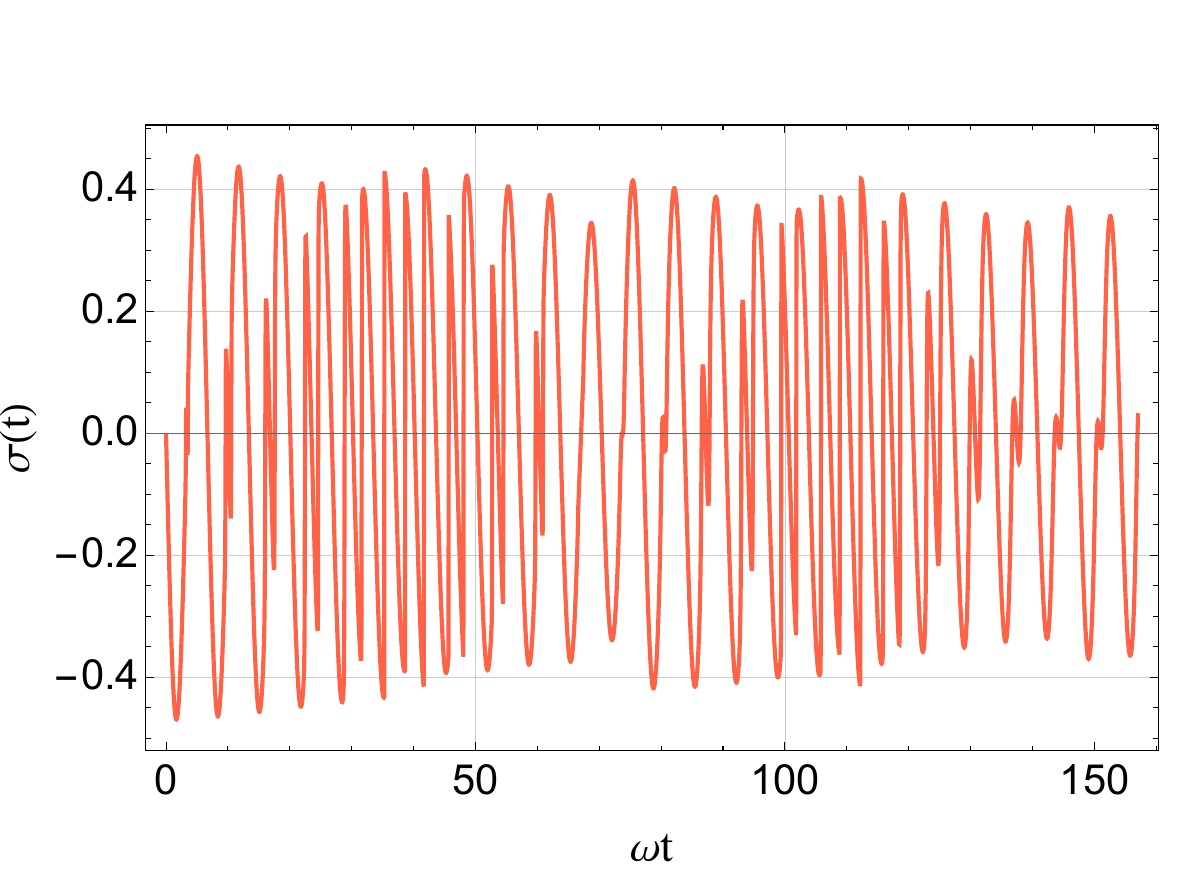}
\caption{Time evolution of the \textbf{(a)} trace distance $D(\rho_1,\rho_2)$ and \textbf{\textbf{(b)}} the trace distance rate of change $\sigma(t,\rho_{1,2}(0))$ as defined, respectively, in ~\eqref{eq:tracedistance} and ~\eqref{eq:tracedistancerate}. The simulations make use of states $\rho_1(0) = | +\rangle \langle +|$ and $\rho_2(0) = | - \rangle \langle - |$ and physical parameters $\omega = 1$, $J = 0.2$,  $p = 5$, $\Gamma=10^{-3}$, $n = 3$, $\omega_D= 1/3$, $\lambda= 3\times10^{-3}$, $\omega t_f = 50 \pi$. All parameters are expressed in dimensionless units by fixing the system frequency $\omega$ as the reference scale.}
\label{fig:evolution__nonmarkov}
\end{figure}
The quantities plotted in Fig.~\ref{fig:evolution__nonmarkov} show a non-monotonic behavior with oscillations exhibiting a
progressive damping due to the dissipative effects of the bath on the two-level system, and a further modulation that can be traced back to the drive contribution. Accordingly, the system undergoes a manifest non-Markovian dynamics due to the back-action induced by the harmonic oscillator.  
{The presence of memory effects, witnessed by  the non-monotonic behavior of $D(t)$ and $\sigma(t)$, will also play an important role in the thermodynamic characterization of the system, which is addressed in the next paragraph.}

\begin{figure}[t!]
\makebox[\columnwidth][l]{\hspace{0.01\columnwidth}\textbf{(a)}}\\
        \includegraphics[width=\columnwidth]{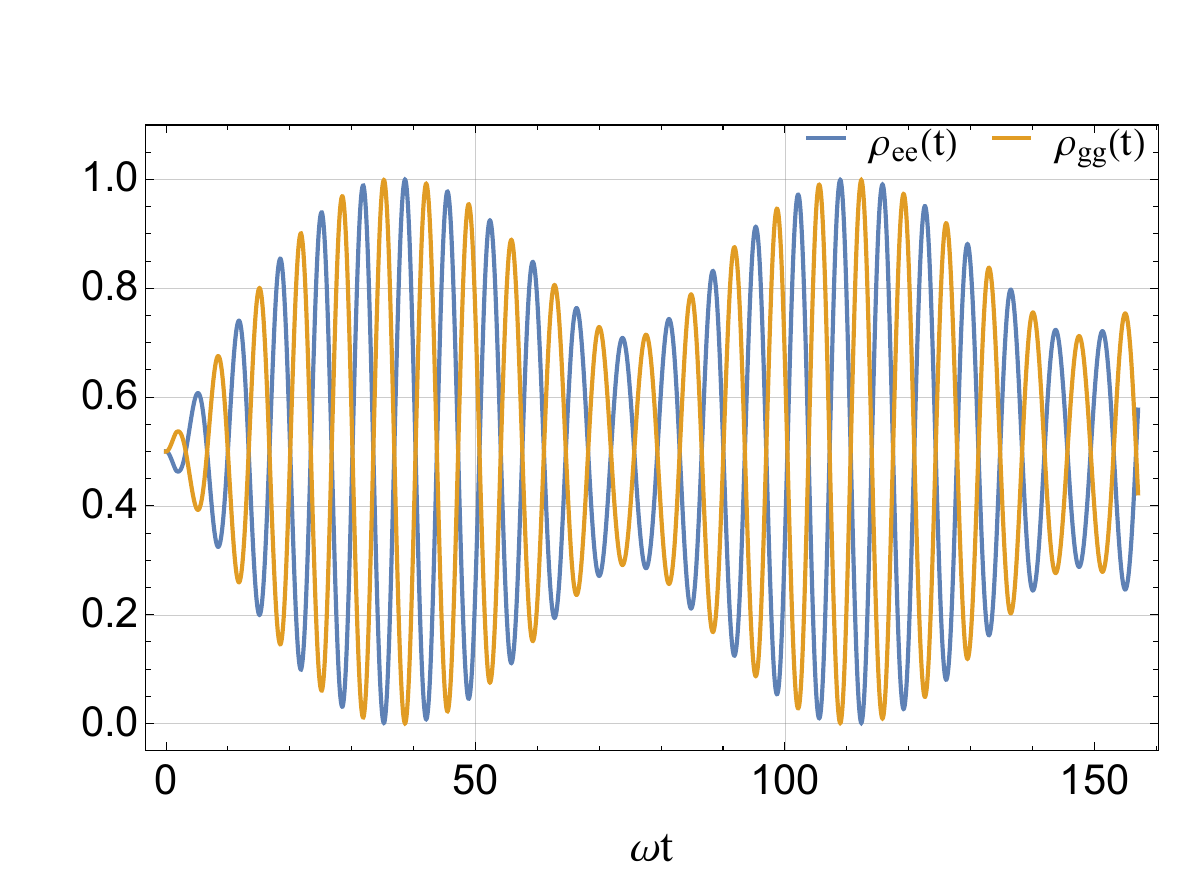}
\makebox[\columnwidth][l]{\hspace{0.01\columnwidth}\textbf{(b)}}\\
        \includegraphics[width=\columnwidth]{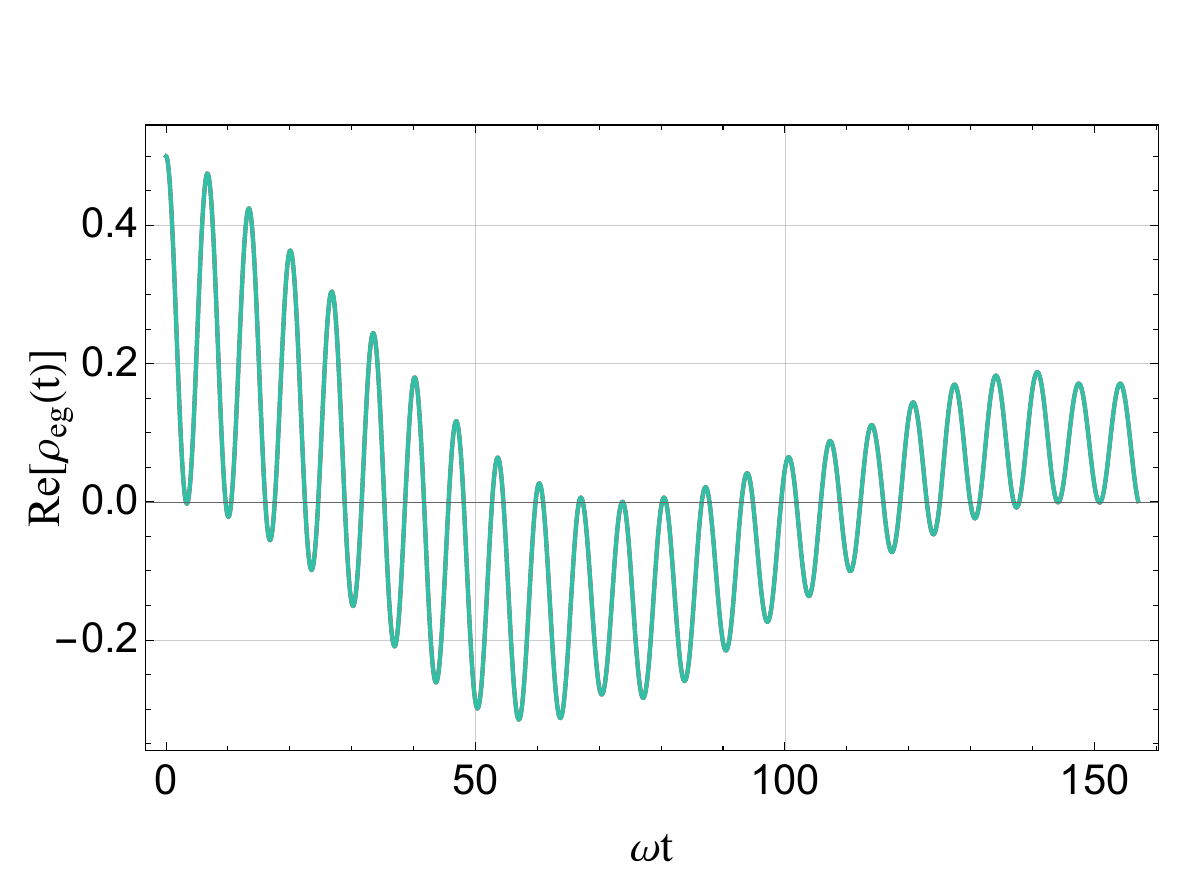}

\caption{Temporal behavior of the \textbf{(a)} populations and \textbf{(b)} coherences of the system's density matrix plotted against the dimensionless time $\omega t$. We only report the real part of coherences ($\text{Re}[\rho_{eg}]=\text{Re}[\rho_{ge}]$), since the imaginary one is quantitatively negligible. 
In all the simulations, if not otherwise specified, we used the same parameters as in Fig.~\ref{fig:evolution__nonmarkov}.}
\label{fig:evolution}
\end{figure}
We now turn to the analysis of the dynamics of the qubit while interacting with the other subsystems. {In doing this,  we assume the resonance condition $ \omega_S = \omega_{ho} \equiv \omega$. Moreover, unless otherwise specified, the simulations that follow are obtained assuming $\rho_S(0)= | +\rangle \langle + |$ as initial state of the qubit.} For an oscillator initially prepared in a number state with $p$ excitations, $\rho_{E}(0) = |p \rangle \langle p|$, we can solve numerically Eq.~\eqref{eq:ME}, where we replaced the $\gamma_i(t)$ rates with the more general quantities $\Gamma_1(t) = \gamma_1(t) + \Gamma (n+1)$, $\Gamma_2(t) = \gamma_2(t)  + \Gamma n$ and $\Gamma_3(t) = \gamma_3(t)$, which include the bath features.
{The outcomes of the resulting equations are shown in Fig.~\ref{fig:evolution}, where we see that populations and coherences exhibit a beat pattern, typical of a Jaynes-Cummings-like interaction. This behavior originates from the interference of oscillations induced by the harmonic oscillator at nearby Rabi frequencies $\Omega_{p}$ and $\Omega_{p+1}$.}

Moreover, we notice that the system does not seem to reach any steady state within the considered time scale, with populations that oscillate indefinitely and coherences that do not decay.
Such a behavior can be linked to a non-Markovian dynamics for the system, in complete agreement with our prediction from the BLP measure in Fig.~\ref{fig:evolution__nonmarkov}.

A comparison with the special cases in Appendix~\ref{appendix2} unveils how such a feature derives from the interactions of the qubit with the harmonic oscillator. The inclusion of the bath, instead, produces further effects of dissipation and decoherence in the dynamics of the two-level system. Indeed, the inversion of populations between the qubit's states $| e \rangle$ and  $| g \rangle$, described from the plot in Fig.~\ref{fig:evolution} {\bf (a)}, is not complete as it would be in the no-bath case. The reason for which this phenomenon takes place is that the reservoir tends to steal excitations (and energy) from the system. 

The driving term, instead, gives rise only to minor alterations, which come from the superposition of an additional signal, of frequency $\omega_D$, on $\rho_{ee}(t)$, $\rho_{gg}(t)$ and  $\text{Re}[\rho_{ge}]$.

\subsection{Thermodynamics}\label{section:results_thermodynamics}
Let us now switch to an analysis of the thermodynamics quantities that characterize the two-level system while interacting with its environment and being driven from the external field.
\subsubsection{Work and heat}\label{section:results_heat&work}

We start by adapting Eq.~\eqref{eq:derwork} and Eq.~\eqref{eq:derheat} to our case, leading to the two differential equations
\begin{align}
	\dot{W}(t) &= - \lambda \omega_D \sin (\omega_D  t)[ \rho_{eg}(t) + \rho_{ge}(t)] \notag \\
	&\quad + i \lambda \omega_D \cos (\omega_D  t)[ \rho_{eg}(t) - \rho_{ge}(t)], \label{eq:drive1_work} \\
    \dot{Q}(t) &=2[ \Gamma_2(t) \rho_{gg}(t) - \Gamma_1(t) \rho_{ee}(t)] \label{eq:envheat}.
\end{align}
The solutions of these equations are shown, respectively, in Fig.~\ref{fig:work&heat} \textbf{(a)} and Fig.~\ref{fig:work&heat} \textbf{(b)}.
From here, we notice that the work only takes negative values, that, in light of the adopted convention, means that work is executed on the system. In this case, the qubit is subject to an external drive -- the rotating magnetic field 
-- which represents the physical means by which work is actually performed. However, the two-level system is also interacting with the harmonic oscillator, and thus the curve profile experiences fluctuations, ultimately heading to a superposition of oscillations.

\begin{figure}[t!]
\makebox[\columnwidth][l]{\hspace{0.01\columnwidth}\textbf{(a)}}\\
        \includegraphics[width=\columnwidth]{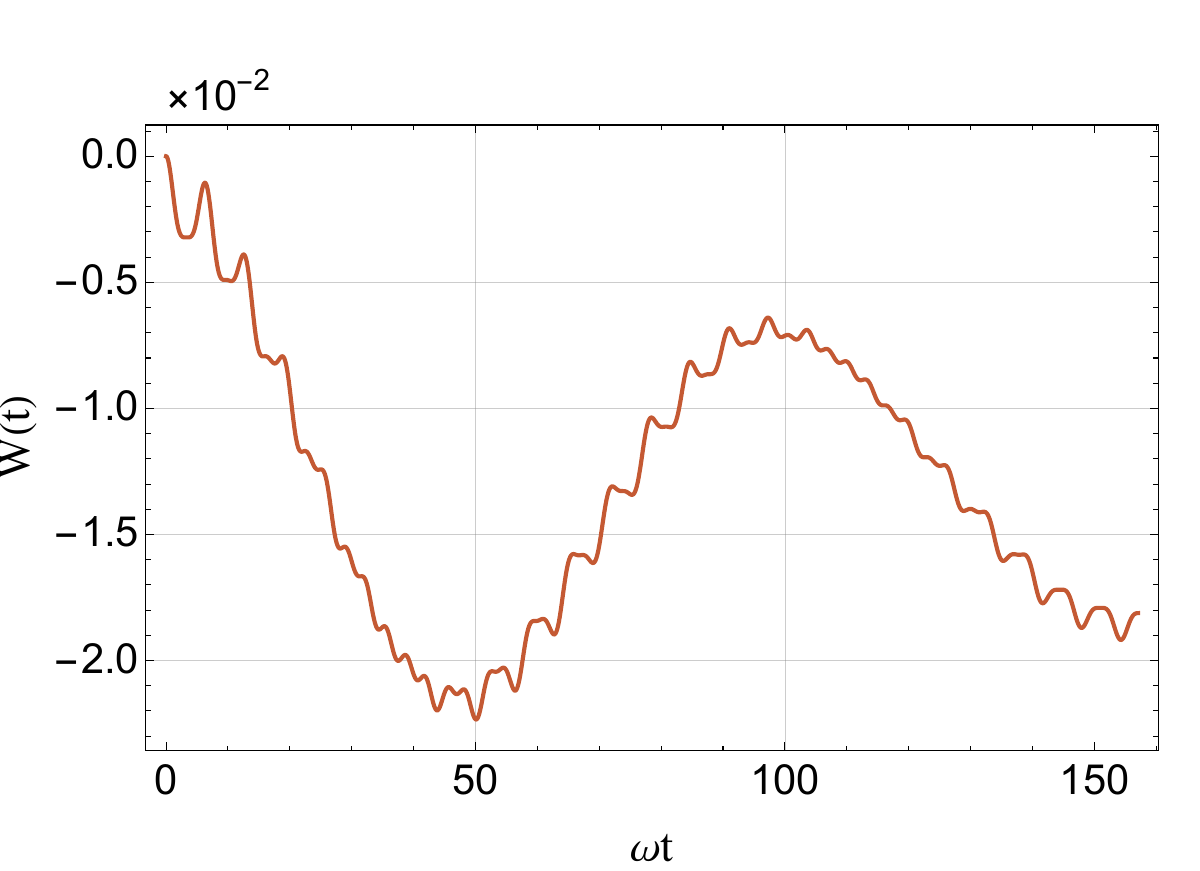}\\
\makebox[\columnwidth][l]{\hspace{0.01\columnwidth}\textbf{(b)}}\\
        \includegraphics[width=\columnwidth]{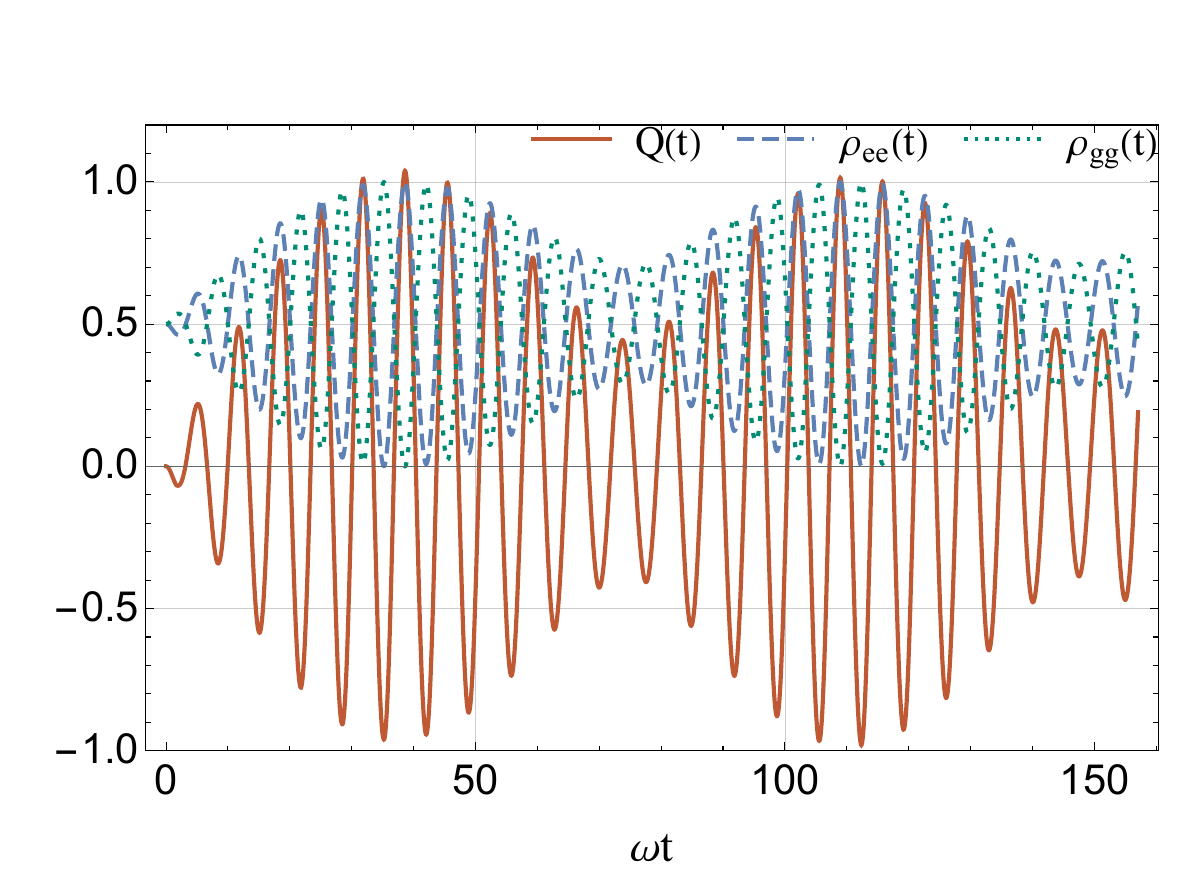}\\
    \caption{Panel {\bf (a)}: Temporal behavior of work done on the system, plotted against the dimensionless time $\omega t$.  
    Panel \textbf{(b)}: We show the dynamics of the heat $Q(t)$ exchanged between the qubit and the environment for the same situation displayed in panel {\bf (a)}, along with the system populations for comparison.
   In all the simulations, if not otherwise specified, we used the same parameters as in Fig.~\ref{fig:evolution__nonmarkov}.}
	\label{fig:work&heat}
\end{figure}

The interaction with the reservoir damps the oscillations of $W(t)$, which, in the long-time limit, decay to a regime where only small fluctuations remain.

Moving to the analysis of $Q(t)$, we notice that the heat profile follows from the behavior of the populations and the dynamical exchange of excitations. In particular, the energy transfer that occurs between the two parts determines the sign of $Q(t)$.
From Fig.~\ref{fig:work&heat} {\bf (b)}, we deduce that, when $\rho_{gg}(t)$ reaches its maximum, the probability of occupying the ground state approaches one. Therefore, the qubit is releasing excitations and energy, which are absorbed from the oscillator. Correspondingly, the heat takes negative values.
On the other hand, when $\rho_{ee}(t)$ attains its peak, the probability of absorbing excitations from the surrounding grows too, and we notice positive peaks in the heat landscape.
Moreover, the work scale appears small compared to the one of the heat exchanged. This is reasonable considering the weak intensity of the drive used to proceed with simulations. The profile of internal energy $U(t)$ will be, then, qualitatively identical to the heat with small quantitative modifications.

\subsubsection{Entropy production and entropy production rate}\label{section:results_entropyproduction}

\begin{figure*}[t!]
\centering
\makebox[\textwidth]{%
\makebox[0.33\textwidth][l]{\hspace{0.02\textwidth}\bf (a)}%
\makebox[0.32\textwidth][l]{\hspace{0.02\textwidth}\bf (b)}%
\makebox[0.33\textwidth][l]{\hspace{0.02\textwidth}\bf (c)}%
}
\includegraphics[width=0.33\textwidth]{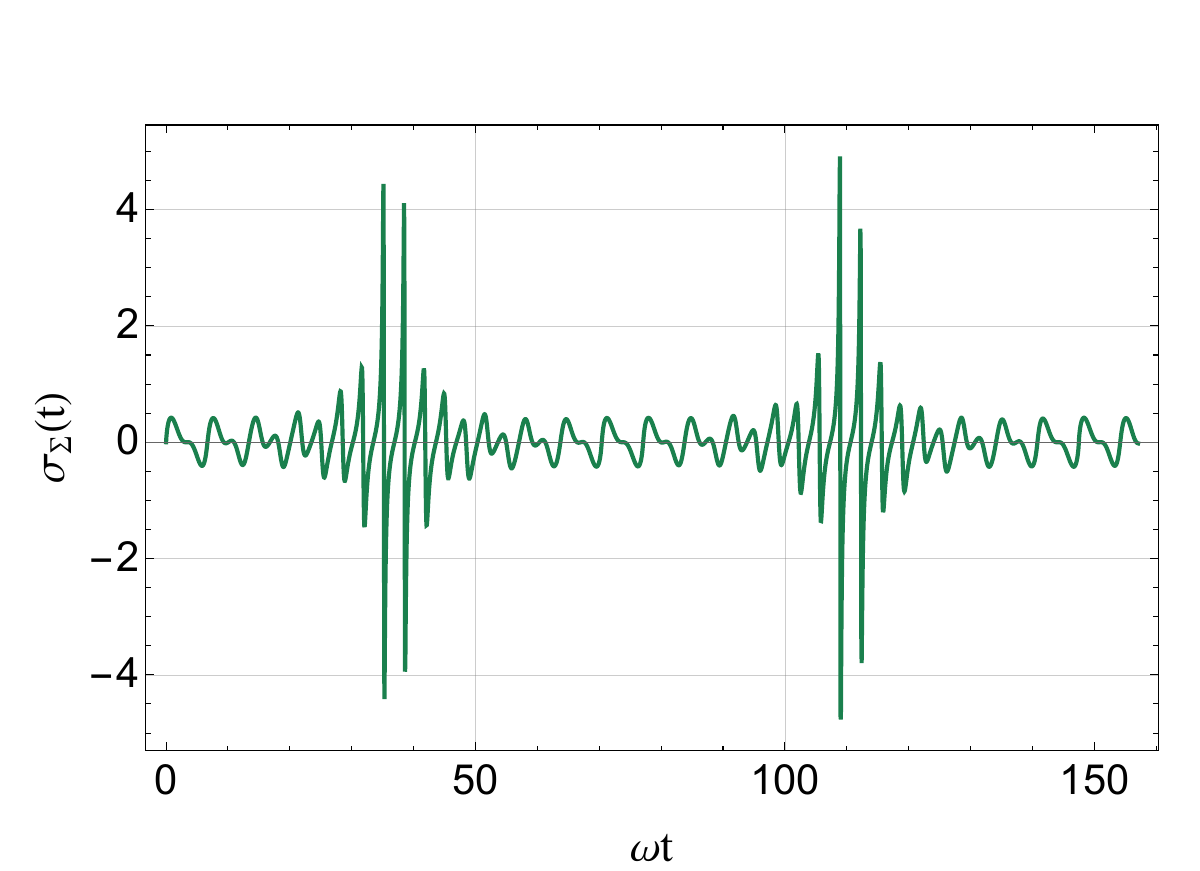}\hfill
\includegraphics[width=0.32\textwidth]{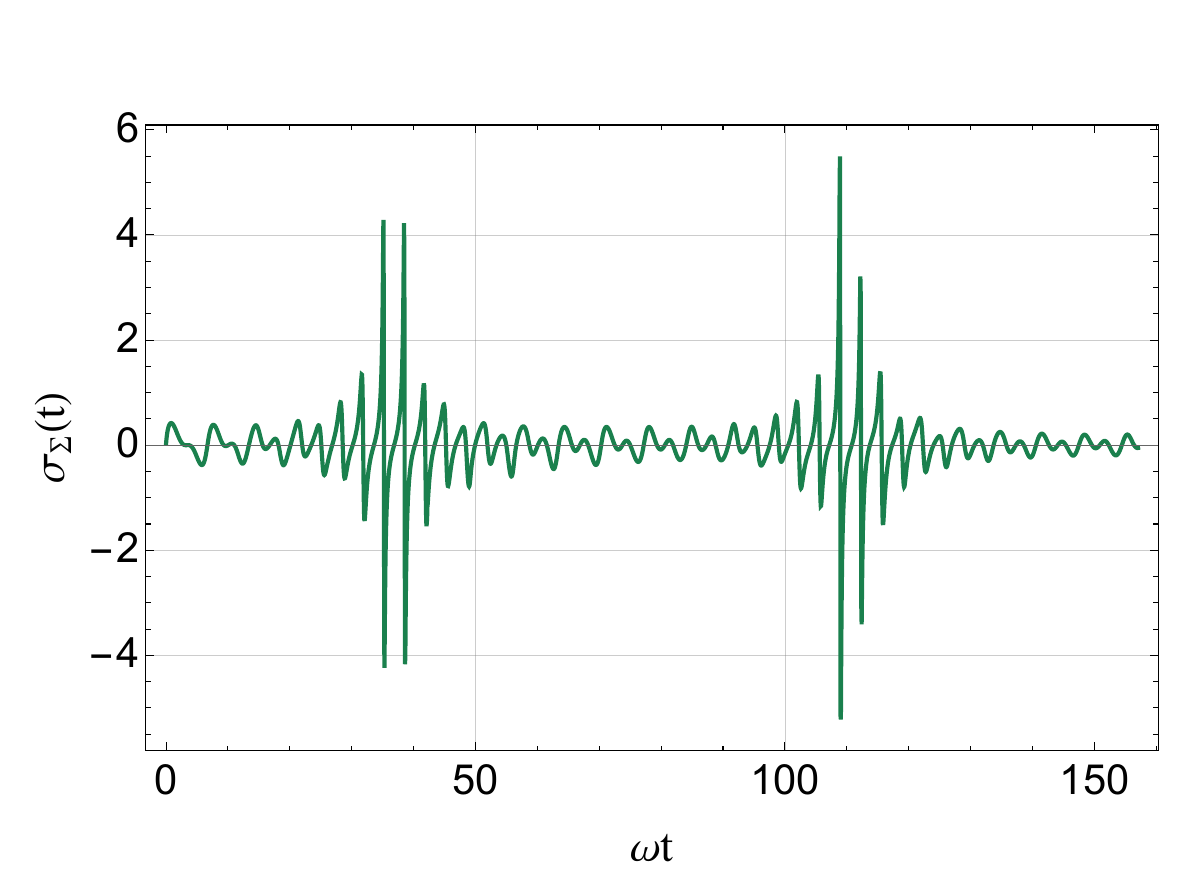}\hfill
\includegraphics[width=0.33\textwidth]{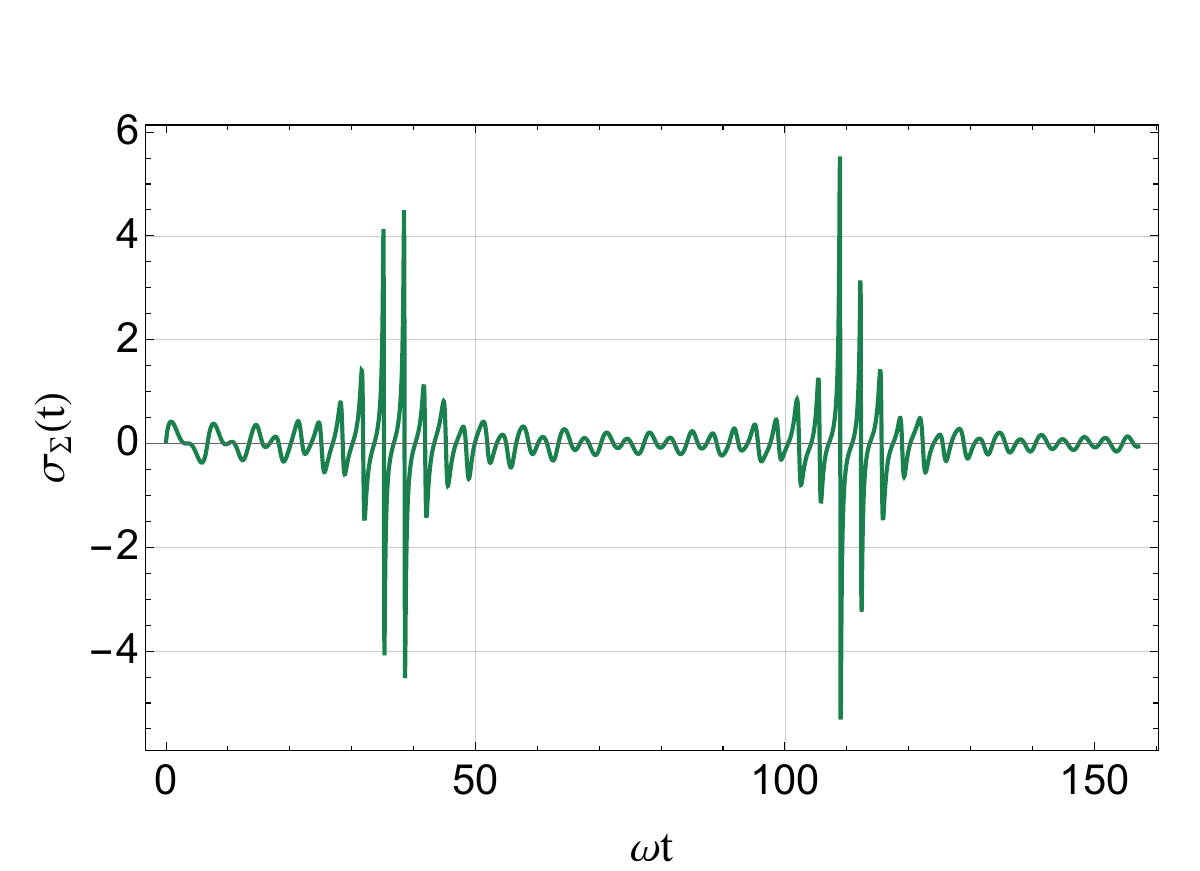}
\caption{Time evolution of the entropy production rate $\sigma_\Sigma(t)$ for three different dynamical configurations.
In panel \textbf{(a)} we set $\Gamma=\lambda=0$; in panel \textbf{(b)} $\Gamma=10^{-3}$, $\lambda=0$; and in panel \textbf{(c)} $\Gamma=10^{-3}$, $\lambda=3\times10^{-3}$.
In all cases, $\omega = 1$, $J = 0.2$,  $p = 5$, $n = 3$, $\omega_D= 1/3$,  $\omega t_f = 50 \pi$. All parameters are expressed in dimensionless units by fixing the system frequency $\omega$ as the reference scale.}
\label{fig:thermoentropy}
\end{figure*}

\begin{figure*}[t!]
\centering
\makebox[\textwidth]{%
\makebox[0.33\textwidth][l]{\hspace{0.02\textwidth}\bf (a)}%
\makebox[0.32\textwidth][l]{\hspace{0.02\textwidth}\bf (b)}%
\makebox[0.33\textwidth][l]{\hspace{0.02\textwidth}\bf (c)}%
}
\includegraphics[width=0.33\textwidth]{media/new_entropyrate_padding.pdf}\hfill
\includegraphics[width=0.32\textwidth]{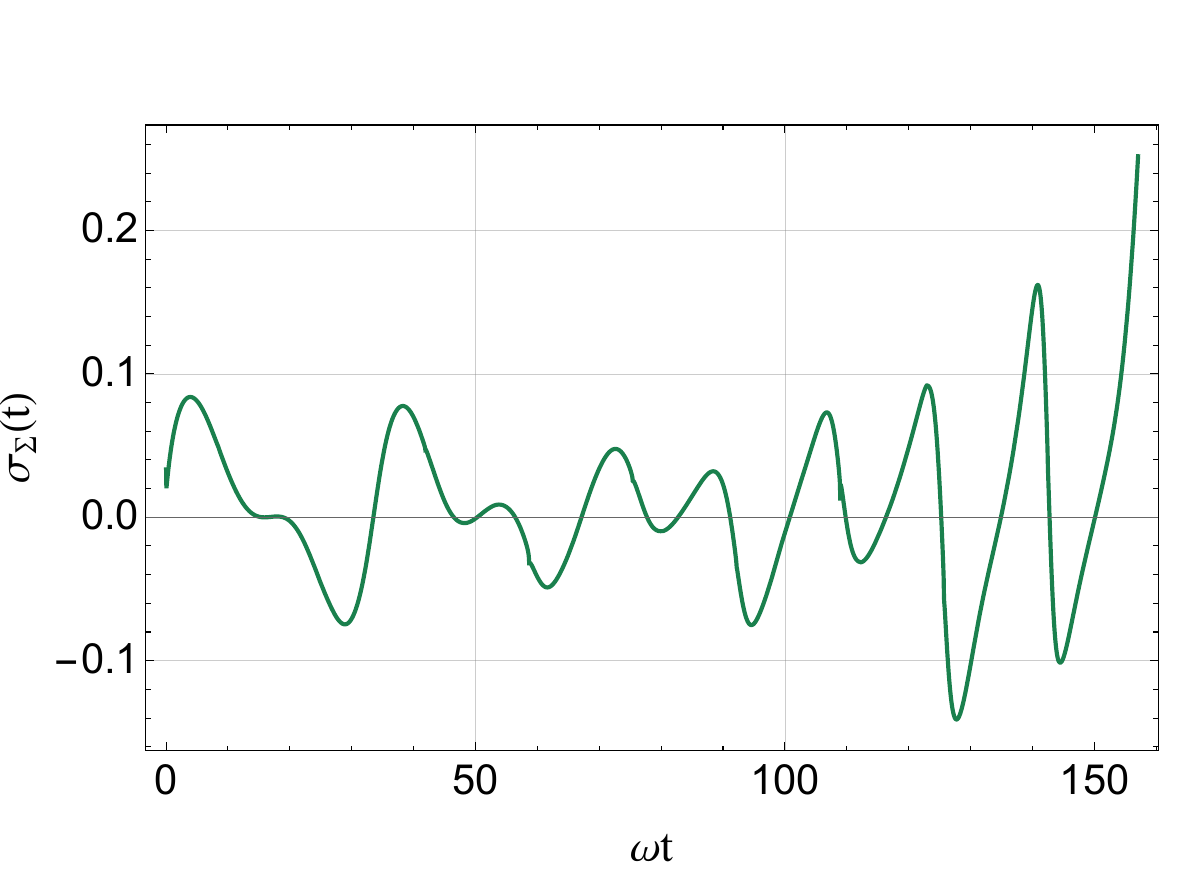}\hfill
\includegraphics[width=0.33\textwidth]{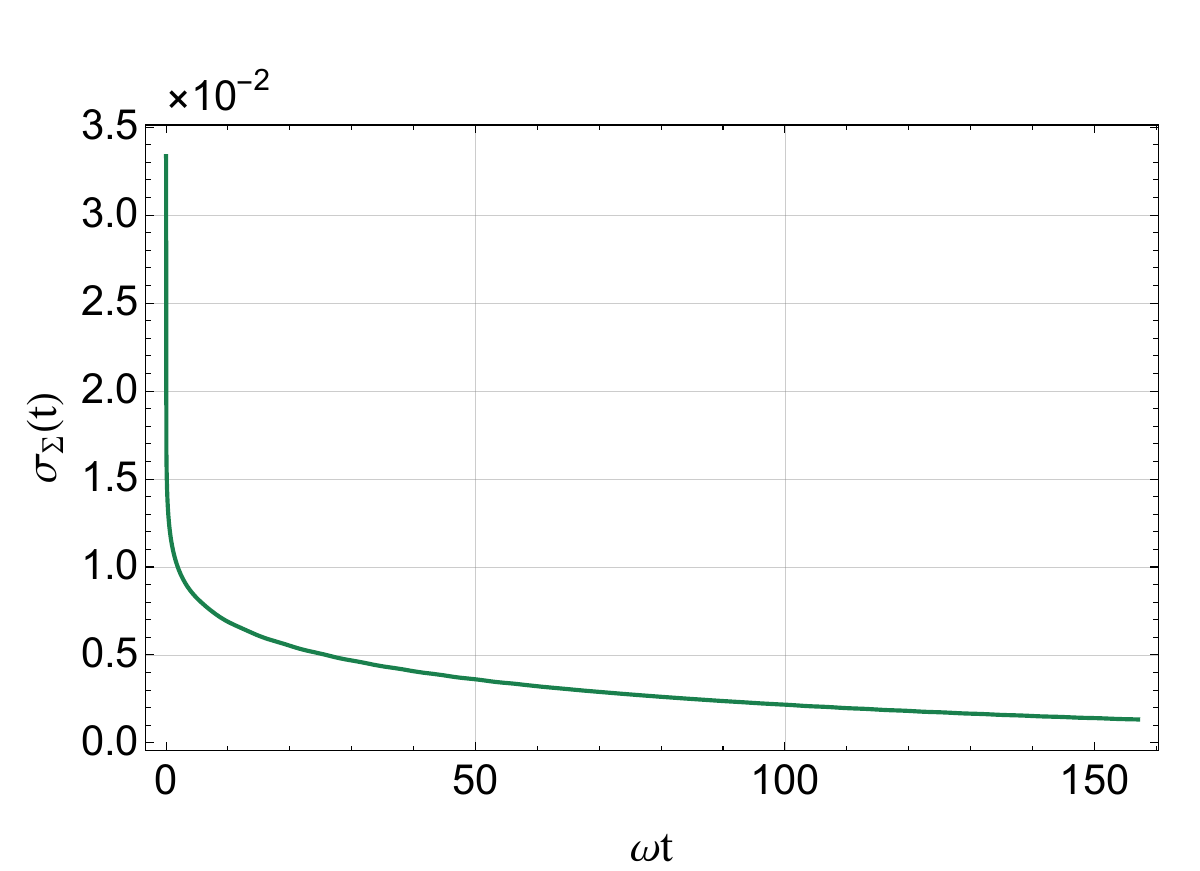}
\caption{Time evolution of entropy production rate $\sigma_\Sigma(t)$ for various choices of the system-oscillator coupling strength. We have taken $J=0.2, 0.04$ and $0$ in panel \textbf{(a)}, \textbf{(b)} and \textbf{(c)}, respectively. In all cases, $\omega = 1$, $p = 5$, $\Gamma=10^{-3}$, $n = 3$, $\omega_D= 1/3$, $\lambda= 3\times10^{-3}$, $\omega t_f = 50 \pi$. All parameters are expressed in dimensionless units by fixing the system frequency $\omega$ as the reference scale.}
\label{fig:J_rates}
\end{figure*}

We proceed similarly with the analysis of the entropy production $\Sigma(t)$, which takes into account the amount of entropy generated by the qubit during the interaction with the environment, and its rate of change, $\sigma_{\Sigma}(t)$.
The entropy production $\Sigma (t)$ can be cast as\textcolor{black}{
\begin{equation}
	\Sigma (t)= S(t) -\beta (t) Q(t),
	\label{eq:entropyprod}
\end{equation}
where $\Delta S(t) = S(t) - S(0)$ simplifies to $ S(t)$ given that the initial state of the qubit is a pure state and $S(0)=0$.
}

However, we need to consider that the currently studied setup is made up of a two-level system in contact with a \textit{single} harmonic oscillator and a bosonic bath.
We then need to distinguish the features of the two subsystems constituting the qubit's environment. Under these conditions, the entropy production $\Sigma (t)$ can be rewritten as
\textcolor{black}{
\begin{equation}
    \Sigma(t)  = S(t) -  \beta_{ho} (t) Q_{ho}(t) -  \beta_{B} (t) Q_{B}(t).
    \label{eq:entropyprod_ho+bath}
\end{equation}}
In Eq.~\eqref{eq:entropyprod_ho+bath}, the heat is decomposed as a sum of two contributions $Q(t)=Q_{ho}(t)+Q_B(t)$, that can be derived from $\dot{Q}_{ho}(t) =2[ \gamma_2(t) \rho_{gg}(t) - \gamma_1(t) \rho_{ee}(t)]$ and $\dot{Q}_{B}(t) =2[ \Gamma n \ \rho_{gg}(t) - \Gamma (n+1) \rho_{ee}(t)]$. They respectively represent the rate of heat exchanged between the qubit and the harmonic oscillator only, and the one exchanged with the reservoir. 
We associate an inverse temperature $\beta_B$ to the latter by using the assumptions of bosonic bath and the standard simplification of a reservoir containing an average number of excitations $n$ and made up of a set of equal harmonic oscillators with frequency $\omega$ (for the sake of argument, we maintain the resonance condition invoked previously)\cite{Breuer}
\begin{equation}
    \beta_B =  \frac{1}{\omega} \ln \bigg( \frac{n+1}{n} \bigg).
\end{equation}
Regarding, instead, the harmonic oscillator, one way to associate with it an effective temperature is by assuming our system instantaneously in equilibrium with the environment. 
To begin with, we build the state $\rho_S^{\text{diag}}(t) = \mathrm{diag} (\rho_{ee}(t),\rho_{gg}(t))$, obtained by removing the off-diagonal elements of the density matrix $\rho_S(t)$.
Then, we associate with $\rho_S^{\text{diag}}(t)$ a fictitious thermal state of inverse temperature $\beta_{ho} (t)$, writing the relation
\begin{equation}
	\rho_S^{\text{diag}}(t) = 
	\begin{pmatrix}
		e^{-\beta_{ho}(t)\omega/2}/Z(t)  & 0\\
		0 & e^{\beta_{ho}(t)\omega/2}/Z(t)
	\end{pmatrix} = \rho_S^{\text{th}}(t) ,
	\label{eq:instantaneous_thermalstate} 
\end{equation}
where $Z(t)$ is the time-dependent partition function.
This equivalence leads us to the expression
\begin{equation}
\beta_{ho}(t) =- \frac{1}{\omega} \ln \Bigg(  \frac{\rho_{ee}(t)}{\rho_{gg}(t)} \Bigg).
\label{eq:hoinverseT}
\end{equation}
Thus, the quantity found in Eq.~\eqref{eq:hoinverseT} is an effective inverse temperature, formally associated with the qubit. However, it reflects the temperature that an external environment (the harmonic oscillator, in this case) would have if it were described by a thermal state and in equilibrium with the system.

In addition, we also need to stress here that the factor $\beta_{ho}(t)$ describes the inverse temperature of the system (and by extension, the harmonic oscillator) when we \textit{force} its description through a thermal, equilibrium state.
Indeed, because of the inversion of population that occurs in the dynamics of the qubit, $\beta_{ho}(t)$ can also assume negative values. This can also be interpreted as a warning of non-equilibrium dynamics: the system in contact with the harmonic oscillator cannot be described by a \textit{well-defined} thermal state, in the sense that no strictly-positive temperature can be associated with it.

{In Fig.~\ref{fig:thermoentropy}, we show the simulations of the rate $\sigma_{\Sigma} (t) = d\Sigma(t)/dt$ for three different cases. Indeed, while entropy production constitutes the central thermodynamic quantity in our analysis, the rate provides a more direct and sensitive characterization of the dynamics. Moreover, being the integrand of the total entropy production, $\sigma_{\Sigma} (t)$ captures the same physical content as ${\Sigma} (t)$
.}

{Thus, in Fig.~\ref{fig:thermoentropy} {\bf (a)} we report the rate of entropy produced by the qubit while interacting solely with the oscillator ($\Gamma = \lambda = 0$). A comparison with the plots in Fig. ~\ref{fig:evolution} shows that the rate attains its largest value when the probability of finding the qubit in the pure states $| e\rangle \langle e|$ or  $| g\rangle \langle g|$ is the highest. Correspondingly, we witness the maximum production of entropy by the two-level system.} 
These states are the furthest from equilibrium, {showing that irreversibility is generated when the system is driven away from equilibrium and confirming the relevance}
of the entropy production and its rate in characterizing non-equilibrium processes.
{The scale of $\sigma_{\Sigma} (t)$ significantly decreases in the central and extreme regions, signaling that $\Sigma (t)$ reaches its minimum}. The reason for this trend is due, once again, to the behavior of populations and coherences. 
{In view of this, we also notice that the rate takes both positive and negative values. This feature confirms our expectations, being a typical signature of non-Markovianity in thermodynamics properties ~\cite{Steve, Marcantoni}.}

One last comment regarding the plots shown in Fig.~\ref{fig:thermoentropy} {\bf (a)} relates to the emergence of a certain overall periodicity in the plots. This apparent "symmetry" echoes the behavior characterizing the dynamics of the elements of $\rho_S (t)$. Such a feature is here preserved, apart from small numerical artifacts that may slightly affect some graphs.
Moving to Fig.~\ref{fig:thermoentropy} {\bf (b)}, we can, instead, observe a sort of breaking of such a periodicity. The reason for this event can be related to the effects of the bath on the dynamics, resulting in larger amount of dissipated energy.
Finally, in Fig.~\ref{fig:thermoentropy} {\bf (c)} we witness the addition of the driving, which produces a modulation of the oscillations that characterize $\sigma_{\Sigma}(t)$, related to the presence of a further, external signal. Indeed, the amplitudes of the waves in the central and side regions of the rate of entropy production plot appear more damped with respect to the previous case.

{The entropy production rate can be used to further investigate the memory effects characterizing our system, and ultimately to witness the emergence of the Markovian limit.}
Indeed, we observe in the panels of Fig.~\ref{fig:J_rates} that, within the same fixed time interval, the part of the $\sigma_{\Sigma}(t)$ curve which enters the negative semi-plane is significantly reduced as the qubit-oscillator coupling $J$ decreases. The limiting case is represented from the plot in Fig.~\ref{fig:J_rates} {\bf(c)}, where $J=0$ and the rate becomes monotonic and rapidly decays to zero. This allows us to identify the harmonic oscillator as the source of the non-Markovianity for the dynamics of our setup. In particular, here we claim that as the coupling between the oscillator and the qubit is weakened, memory effects are gradually downsized. This persuades us to identify the limit $J \to 0$ as the Markovian limit of the dynamics of the analyzed system. 

\section{CONCLUSIONS}\label{section:conclusions}

We have explored many aspects of the analysis, characterization and simulation of the dynamics of a composite open system. We modeled a quantum device made up of a two-level system interacting with a structured environment, including a harmonic oscillator and a bath. The system is also subject to an external driving field, which can be tailored in order to perform different operations on the qubit.

We described the reduced state of the qubit progressively, through the construction of a Lindblad-like master equation.
At this stage, we were able to identify a non-Markovian behavior for the evolution of the two-level system, due to the interaction with a finite-size environment and the mutual influence between system and environment. This aspect was investigated through the use of the BLP non-Markovianity measure~\cite{PhysRevLett.103.210401}.

In addition, we discussed the thermodynamic implications of such a rich dynamics, through key concepts in non-equilibrium quantum thermodynamics, particularly regarding the entropy production rate.

Indeed, the analysis of $\sigma_{\Sigma}$, i.e. the entropy production time derivative, showed important features, mainly confirming the implications of the non-Markovian character of the dynamics on the thermodynamics of the system, consistently with existing literature~\cite{Marcantoni, Steve}.
We were able to prove that the non-Markovian behavior depends on the coupling strength with the oscillator: as the value of this parameter is decreased, the evidences of non-Markovianity become weaker, ultimately disappearing for $J \to 0$, which allows us to recover the Markovian limit.

This study also paves the way for possible extensions and generalizations. A natural step would be the adaptation of this procedure for an $N$-level system, instead of a qubit. This would assess how an increased number of internal degrees of freedom would influence both the overall dynamics and thermodynamics. Additionally, a complete study of quantum correlations and coherence would help identify the regimes in which a classical description becomes convenient, offering a deeper understanding of the quantum-classical description.
\acknowledgments
B.G.B. was supported by the MICIN grant EUR2024-15354.
M.P. acknowledges support from the European Union’s Horizon Europe EIC-Pathfinder
project QuCoM (101046973), the Department for the Economy of Northern Ireland under
the US-Ireland R\&D Partnership Programme, the ``Italian National Quantum Science and Technology Institute (NQSTI)" (PE0000023) - SPOKE 2 through project ASpEQCt.

\bibliography{biblio}

\appendix

\section{Quantum dynamical map and master equation}\label{appendix1}
Here we show a detailed derivation of the master equation for the qubit interacting only with the harmonic oscillator. To this end, we make use of the approach put forward in Ref.~\cite{Andersson15082007}. 

We first derive the quantum dynamical map, which encodes the information about the evolution of the two-level system. We assume the initial state of the compound to be factorized $\rho_{SE}(0) = \rho_S(0) \otimes \rho_E(0)$, with the harmonic oscillator to be initially in a number state with $p$ excitations, $ \rho_E (0) =|p \rangle \langle p|$, and the two-level system in $\rho_{S} (0)= \sum_{i,j=e,g}
\ \rho_{ij}(0)|i \rangle \langle j |$ with $\rho_{ij}=\langle{i}\vert\rho(0)\vert{j}\rangle$ such that $
\sum_{i,j=e,g} \ \rho_{ii}=1$.
If we do not consider the presence of any other subsystem, the compound $\mathcal{S}+E$  evolves unitarily as
    \begin{equation}
        \rho_{SE}(t) = U(t)\rho_{SE}(0)U^{\dagger}(t),
        \label{eq:evolvedjointstate}
    \end{equation}
where the time-evolution operator $U(t)$ is found from the Jaynes-Cummings Hamiltonian. We then remove the degrees of freedom of the harmonic oscillator and derive $ \rho_S(t)$ as
    \begin{equation}
    \begin{aligned}
        \rho_S(t) &= \Tr_{E} \left\{ \rho_{SE}(t)\right\}\equiv\phi_t[\rho_S(0)]= \sum_k  W_{k,p} (t) \rho_S(0)  W^{\dagger}_{k,p} (t), 
        \label{eq:operatorsum} 
    \end{aligned}
    \end{equation}
where  each $ W_{k,p} (t)  = \langle k |U(t)|p \rangle\in\mathcal{H}_S $ is a Kraus operator. The evolved state of the two-level system reads
\begin{equation}
    \rho_S(t) =
    {\cal A}(t)|e \rangle \langle e|+{\cal B}(t) |g \rangle \langle g| + ({\cal C}(t) |e\rangle \langle g| +h.c.), 
    \label{eq:krausevolved_densitymatrix}
\end{equation}
with the coefficients 
\begin{equation}
\begin{aligned}
    {\cal A}(t) &= C^2_{p+1}(t) \rho_{ee}(0) + S^2_{p}(t) \rho_{gg}(0) \\
    {\cal B}(t) &= S^2_{p+1}(t) \rho_{ee}(0) + C^2_{p} (t)\rho_{gg}(0) \\
    {\cal C}(t) &= C_{p+1}(t) C_{p}(t)\rho_{eg}(0).
\end{aligned}
\end{equation}
where $C_{k}(t)=\cos(Jt \sqrt{{k}})$ and $S_{k}(t)=\sin(Jt \sqrt{k})$.
We now proceed by introducing the orthonormal operator basis ${B} = \{ \mathbb{I}/\sqrt{2},\allowbreak\ \sigma_x/\sqrt{2},\allowbreak\  \sigma_y/\sqrt{2},\allowbreak\  \sigma_z/\sqrt{2} \}$ for the two-level system, using which we have
    \begin{gather}
    	\rho_S(t) = \sum_{ij} F_{ij}(t) \Tr \big( \rho_S(0)B_j\big)B_i ,\label{eq:krausmatrixdecomposition} \\
    	\text{with} \  \ F_{ij}(t)=\Tr(B_i \phi_t[B_j]).
    	\label{eq:defmatrixF}
     \end{gather}
This expression can also be cast in matrix form, by identifying $r_j=\Tr(B_j \rho_S)$ and then rewriting Eq.~\eqref{eq:defmatrixF} as
$\phi_t[\rho_S(0)] = [\textbf{F}(t)\textbf{r}(0)]^{T}  \textbf{B}$ with 
\begin{widetext}
\begin{equation}
	\textbf{F}(t) = 
	\begin{pmatrix}
		1  &  0 & 0 & 0 \\
		0 & C_{p+1}(t) C_{p}(t) & 0 & 0\\
		0 & 0 & C_{p+1}(t) C_{p}(t) &  0\\
		C^2_{p+1}(t) - C^2_{p} (t) & 0 & 0 & C^2_{p+1} (t)+ C^2_{p}(t) -1
	\end{pmatrix}.
	\label{eq:matrixF}
\end{equation} 
\end{widetext}
With the form of $\textbf{F}(t)$ at hand, we can differentiate the dynamical map to write
\begin{equation}
	\dot{\rho}_S (t)=  [\dot{\textbf{F}}(t)\textbf{r}(0)]^{T}  \textbf{B}.
	\label{eq:evolutionmasterequation}
\end{equation}
To find an explicit form for the master equation we now set $\dot{\rho} = \Lambda(\rho) = [\textbf{A}(t)\textbf{r}(t)]^{T}  \textbf{B}$ with 
\textbf{A}(t) a matrix of elements $A_{ij}(t)=\Tr \{ B_i \Lambda(B_j) \}$. 
By direct comparison, and assuming $\textbf{F}$(t) to be invertible, we deduce 
\begin{equation}
	 \textbf{A}(t)= \dot{\textbf{F}}(t)(\textbf{F}(t))^{-1}
	\label{eq:defmatrixL}
\end{equation}
with
\begin{equation}
	\textbf{A}(t) = 
	\begin{pmatrix}
		0  &  0 & 0 & 0 \\
		0 & \varepsilon(t) & 0 & 0\\
		0 & 0 & \varepsilon(t) &  0\\
		\nu(t)& 0 & 0 & \mu(t)
	\end{pmatrix}.
\end{equation}
It is more convenient to change basis and work with the matrix $\textbf{A}'(t)$ with elements $A'_{rs}(t)=\sum_{ij} A_{ij}(t) \Tr(B_sB_jB_r B_{i})$ to get 

\begingroup
\setlength{\arraycolsep}{4pt} 
\begin{equation}
	\textbf{A}'(t)= 
	\begin{pmatrix}
		\varepsilon (t)+ \tfrac{\mu(t)}{2}  &  0 & 0 &  \tfrac{\nu(t)}{2} \\
		0 & - \tfrac{\mu(t)}{2} &   i \tfrac{\nu(t)}{2} & 0\\
		0 & - i \tfrac{\nu(t)}{2}& - \tfrac{\mu(t)}{2} &  0\\
		\tfrac{\nu(t)}{2} & 0 & 0 &	-\varepsilon(t) + \tfrac{\mu(t)}{2}
    \end{pmatrix}.
	\label{eq:defmatrixA'}
\end{equation}
\endgroup
Using the  approach in Refs.~\cite{Breuer,andersson3},  ${\bf A}'(t)$ is instrumental to get a master equation in diagonal form such as
\begin{equation}
\begin{aligned}
	\dot{\rho}_S(t) &=-i\left[ H(t), \rho_S(t) \right]\\ &+ \sum_{i} \gamma_i(t) \left( L_i\rho_S(t) L_i^{\dagger}
    -\frac{1}{2} \left\{L_i^{\dagger} L_i, \rho_S(t)\right\} \right).
	\label{eq:MEmatrixform2}
\end{aligned}
\end{equation}
The rates $\gamma_i(t)$ are found as the eigenvalues of the submatrix $\mathrm{sub}\textbf{A}'(t)$, obtained from Eq.~\eqref{eq:defmatrixA'} by removing its first row and column, connected to $B_0=\mathbb{I}/\sqrt2$. The jump operators $L_i$ are found through the relation
\begin{equation}
	L_i = \sum_j P_{ji} B_j,
    \label{eq:jumpoperatorsmethod}
\end{equation}
with $P_{ji}$ being the elements of a passage matrix made up of the normalized eigenvectors of $\mathrm{sub}\textbf{A}'(t)$.
The Hamiltonian $H(t)$ in Eq.~\eqref{eq:MEmatrixform2} is a Lamb-shift-like term given by
\begin{gather}
	H(t)=\frac{\kappa ^{\dagger}(t)-\kappa(t)}{2i} \quad
	\text{with} \ \ \kappa(t)=\frac{1}{\sqrt{2}}\sum_i A'_{i0}(t) B_i.
	\end{gather}
An explicit calculation leads us to 
\begin{equation}
\begin{aligned}
	 \gamma_1(t)&=-\frac{\mu(t)+\nu(t)}{2},  \\
	 \gamma_2(t)&=\frac{\nu(t)-\mu(t)}{2},  \\
	\gamma_3(t)&=\frac{\mu(t)-2\varepsilon(t)}{2}
    \label{eq:general_rates}
\end{aligned}
\end{equation}
and 
\begin{equation}
	\textbf{P} = 
	\begin{pmatrix}
		i/{\sqrt{2} }  &  -i/{\sqrt{2}} & 0\\
		1/{\sqrt{2}} & 1/{\sqrt{2}} &   0\\
		0 & 0 & 1
	\end{pmatrix}.
	\label{eq:defmatrixP}
\end{equation}
The jump operators $L_i$ can be computed via Eq.~\eqref{eq:jumpoperatorsmethod} and the Lamb-shift Hamiltonian $H(t)$ as
\begin{equation}
\begin{aligned}
	L_1 &= i\sigma_-,\,L_2 = -i\sigma_+,\,L_3 = \sigma_z,\\
\kappa(t) &=\kappa^{\dagger}(t)=\frac{1}{\sqrt2} \left[\left(\varepsilon(t) +\frac{\mu(t)}{2}\right) {\mathbb{I}}+\frac{\nu(t)}{2} {\sigma_z}\right], 
\end{aligned}
\end{equation}
We thus have $H(t) =0$ and can cast the master equation in the form
\begin{align}
	\dot{\rho}_S(t) &=
	\gamma_1(t)\bigg(\sigma_-\rho_S(t)\sigma_+-\frac{1}{2}\big\{ \sigma_+\sigma_-,\rho_S(t)\big\} \bigg) \notag\\
	&+\gamma_2(t)\bigg(\sigma_+\rho_S(t)\sigma_- -\frac{1}{2}\big\{ \sigma_-\sigma_+,\rho_S(t)\big\} \bigg) \notag\\
	&+ \gamma_3(t)\bigg(\sigma_z\rho_S(t)\sigma_z-\rho_S(t) \bigg).
	\label{eq:master_equation}
\end{align}
By using the definitions of $\mu(t), \nu(t) $ and $\varepsilon(t)$ we find the rates in Eq.~\eqref{eq:general_rates} as 
\begin{widetext}
\begin{align}
	\gamma_1 (t)& = \frac{\Omega_{p}\sin(\Omega_{p}t) + \Omega_{p+1}\sin(\Omega_{p+1}t)}
	{2\cos(\Omega_{p}t) + 2\cos(\Omega_{p+1}t)} - \frac{(\Omega_{p}+\Omega_{p+1})\sin[(\Omega_{p}-\Omega_{p+1})t]}
	{4\cos(\Omega_{p}t) + 4\cos(\Omega_{p+1}t)} - \frac{(\Omega_{p}-\Omega_{p+1})\sin[(\Omega_{p}+\Omega_{p+1})t]}
	{4\cos(\Omega_{p}t) + 4\cos(\Omega_{p+1}t)}, \label{eq:rate1}\\
	\gamma_2 (t)&= \frac{\Omega_{p}\sin(\Omega_{p}t) + \Omega_{p+1}\sin(\Omega_{p+1}t)}
	{2\cos(\Omega_{p}t) + 2\cos(\Omega_{p+1}t)} + \frac{(\Omega_{p}+\Omega_{p+1})\sin[(\Omega_{p}-\Omega_{p+1})t]}
	{4\cos(\Omega_{p}t) + 4\cos(\Omega_{p+1}t)} + \frac{(\Omega_{p}-\Omega_{p+1})\sin[(\Omega_{p}+\Omega_{p+1})t]}
	{4\cos(\Omega_{p}t) + 4\cos(\Omega_{p+1}t)},\label{eq:rate2}\\
	\gamma_3 (t)&= -\frac{\big(\Omega_{p+1}\sin(\Omega_{p}t) \big)
		\tan(\Omega_{p}t/2)\tan(\Omega_{p+1}t/2)}
	{4\cos(\Omega_{p}t) + 4\cos(\Omega_{p+1}t)}-\frac{\big(\Omega_{p}\sin(\Omega_{p+1}t) \big)
		\tan(\Omega_{p}t/2)\tan(\Omega_{p+1}t/2)}
	{4\cos(\Omega_{p}t) + 4\cos(\Omega_{p+1}t)},\label{eq:rate3}
\end{align}
\end{widetext}
with $\Omega_{p}=2J \sqrt{p}$ and $\Omega_{p+1}=2J \sqrt{p+1}$.
\section{Special cases}\label{appendix2}
Here, we discuss some special cases in which the effects of the harmonic oscillator and the bath are mutually switched off. 
This allows us to appreciate the influence that each subsystem has on the qubit's features.{ To this end, we fix $\lambda=0$, in order to focus mainly on the interplay between dissipation, decoherence and memory effects.}

\begin{figure*}[t!]
\centering
\makebox[\textwidth]{%
\makebox[0.33\textwidth][l]{\hspace{0.02\textwidth}\bf (a)}%
\makebox[0.32\textwidth][l]{\hspace{0.02\textwidth}\bf (c)}%
\makebox[0.33\textwidth][l]{\hspace{0.02\textwidth}\bf (e)}%
}

\includegraphics[width=0.33\textwidth]{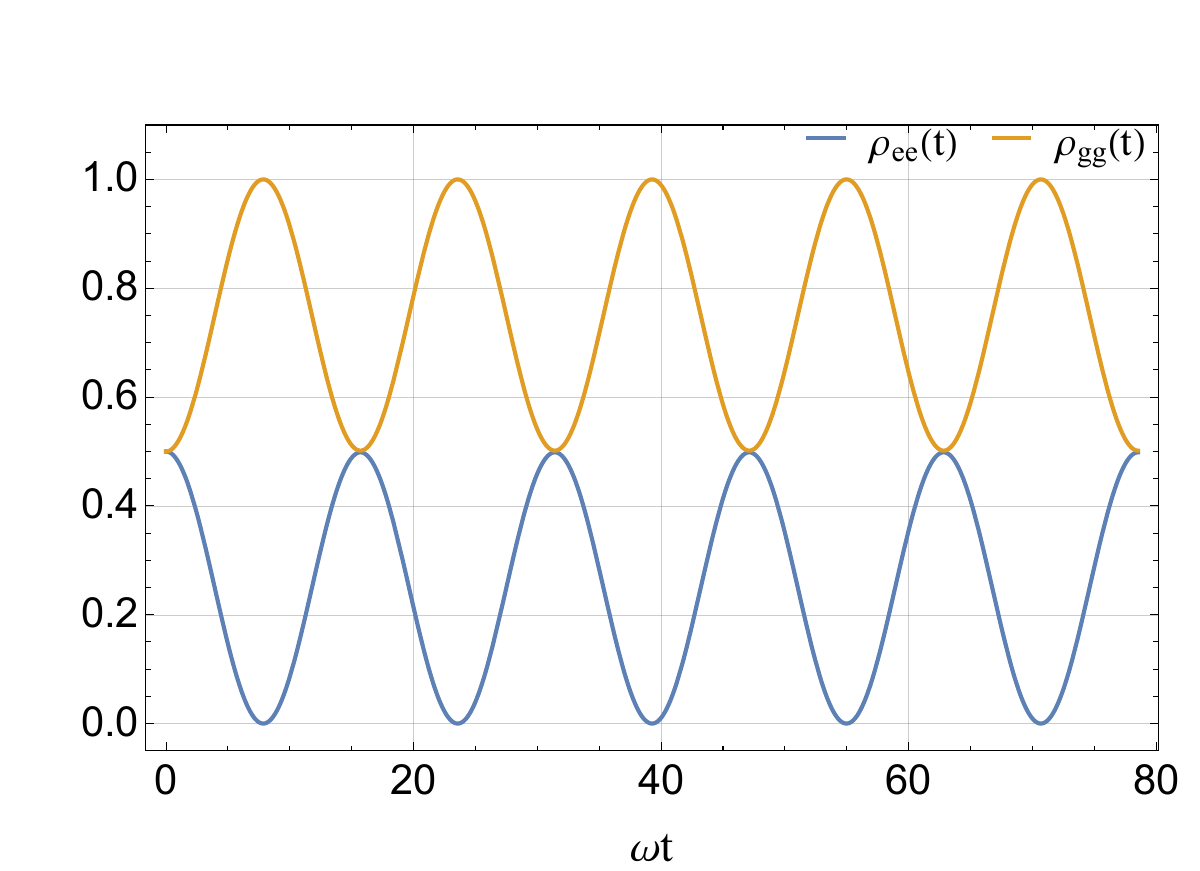}\hfill
\includegraphics[width=0.32\textwidth]{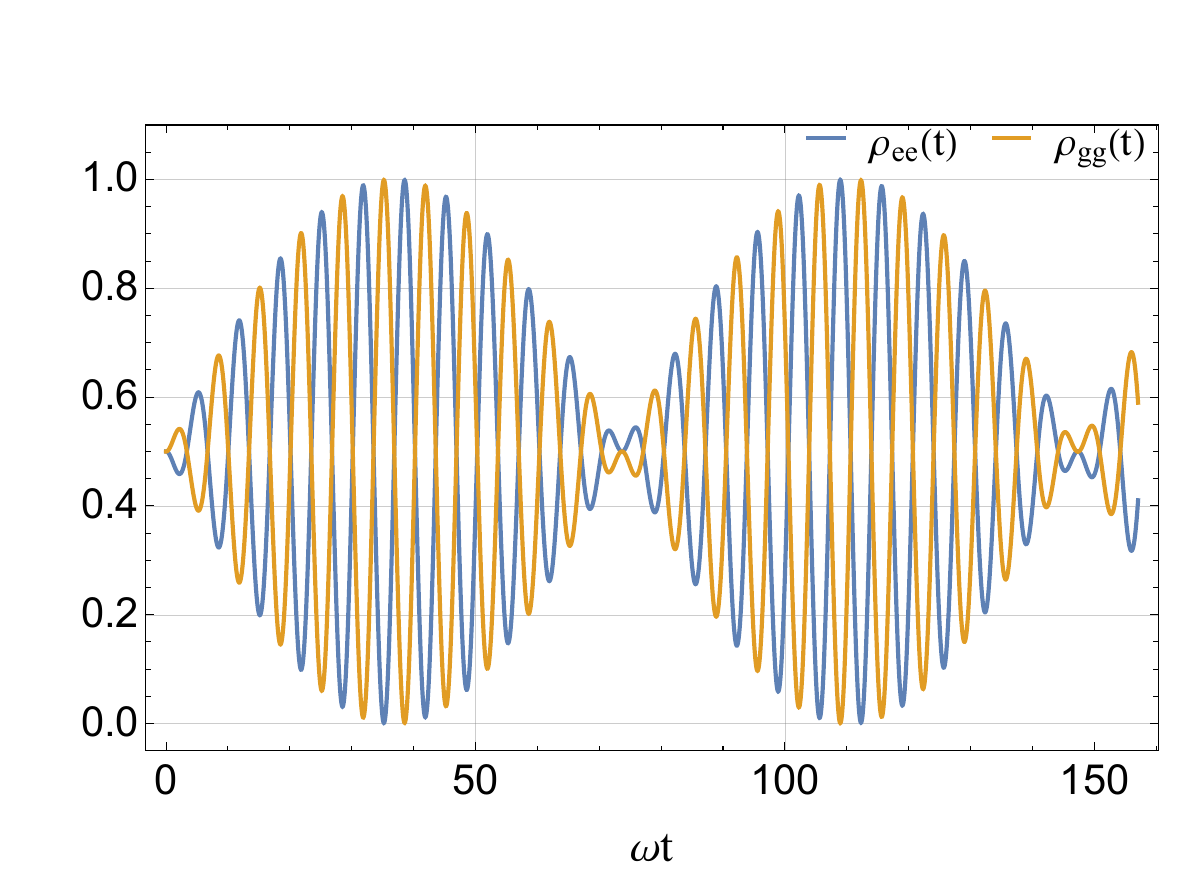}\hfill
\includegraphics[width=0.33\textwidth]{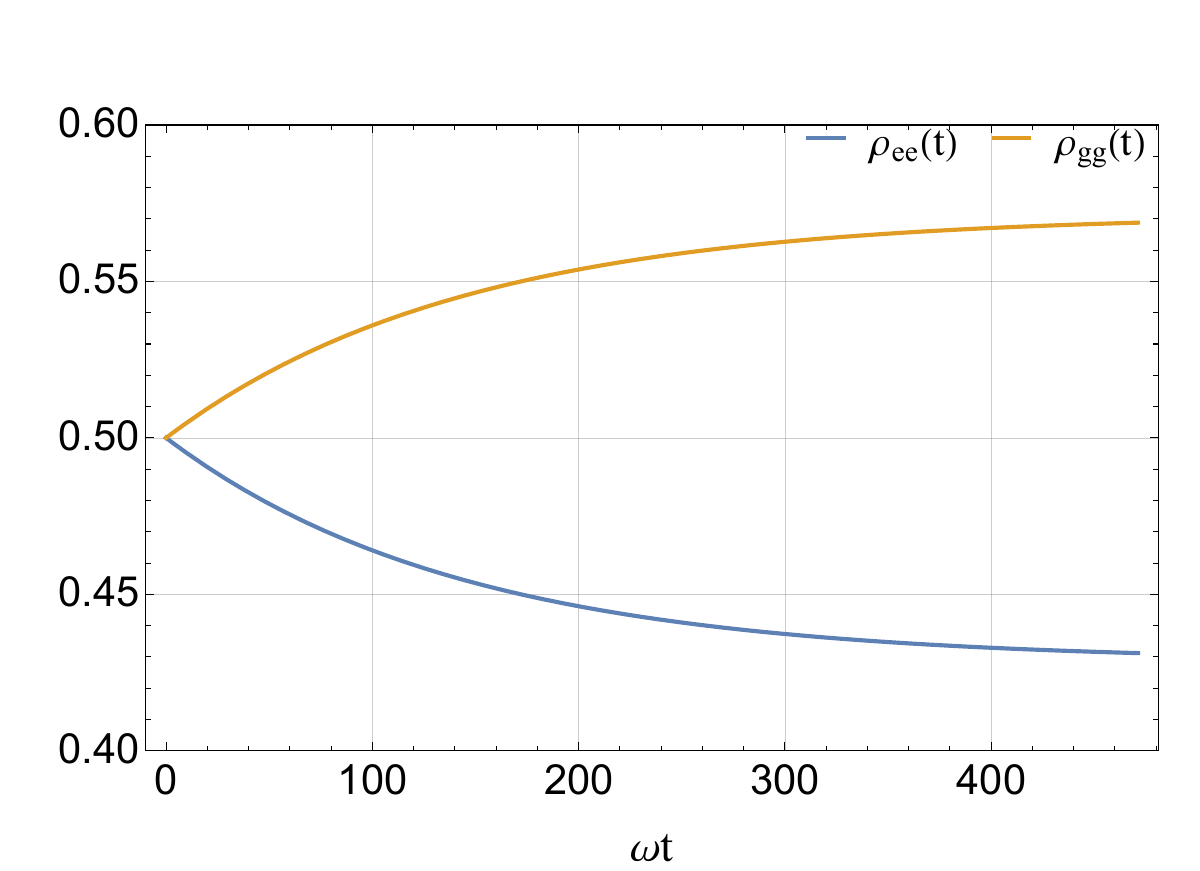}\hfill

\vspace{0.4cm}
\centering
\makebox[\textwidth]{%
\makebox[0.33\textwidth][l]{\hspace{0.02\textwidth}\bf (b)}%
\makebox[0.32\textwidth][l]{\hspace{0.02\textwidth}\bf (d)}%
\makebox[0.33\textwidth][l]{\hspace{0.02\textwidth}\bf (f)}%
}

\includegraphics[width=0.33\textwidth]{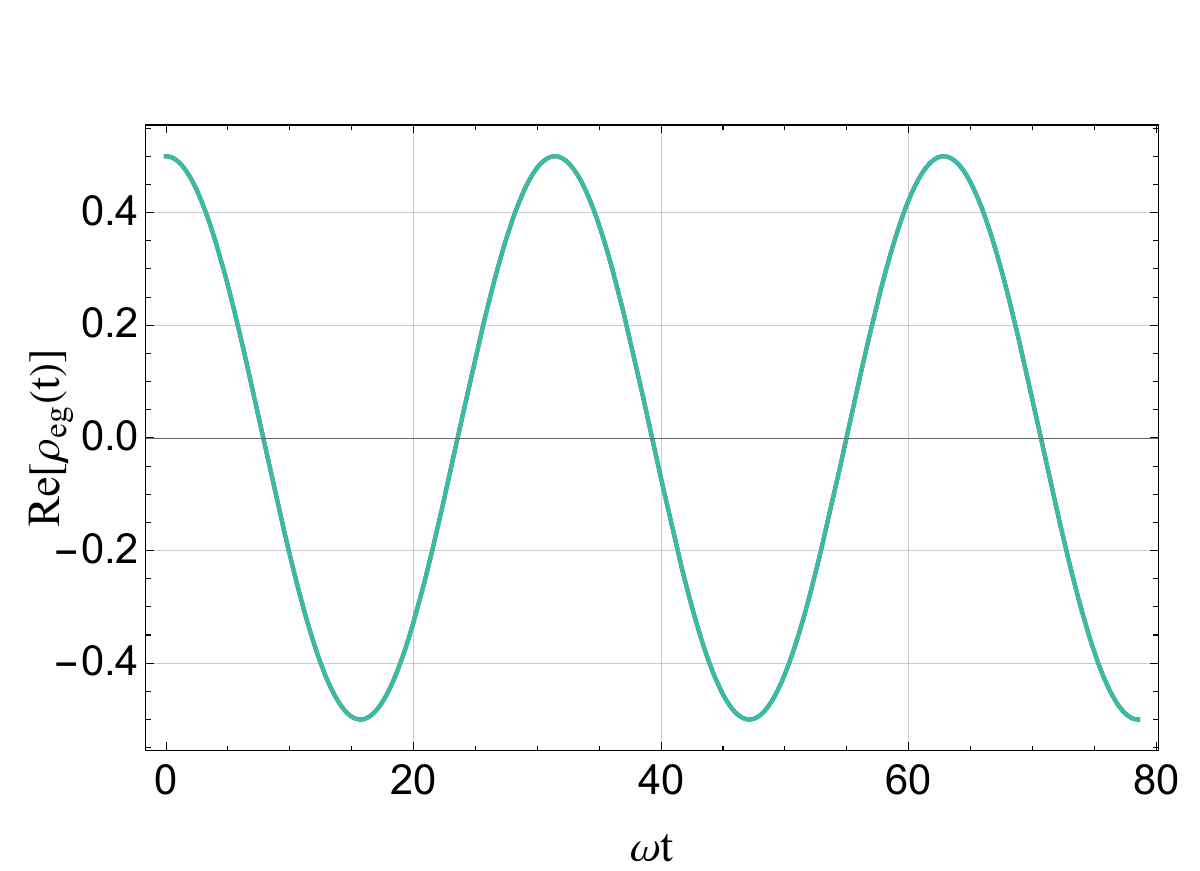}\hfill
\includegraphics[width=0.32\textwidth]{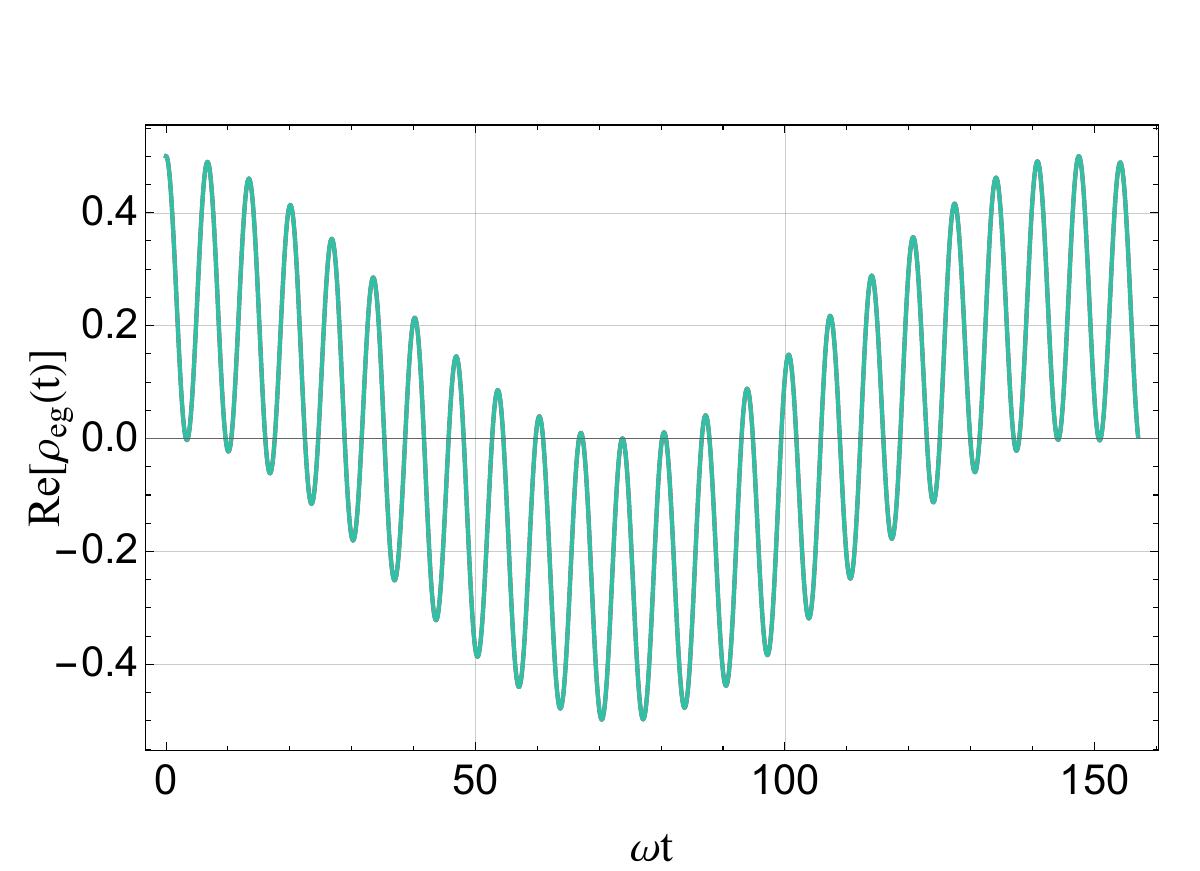}
\includegraphics[width=0.33\textwidth]{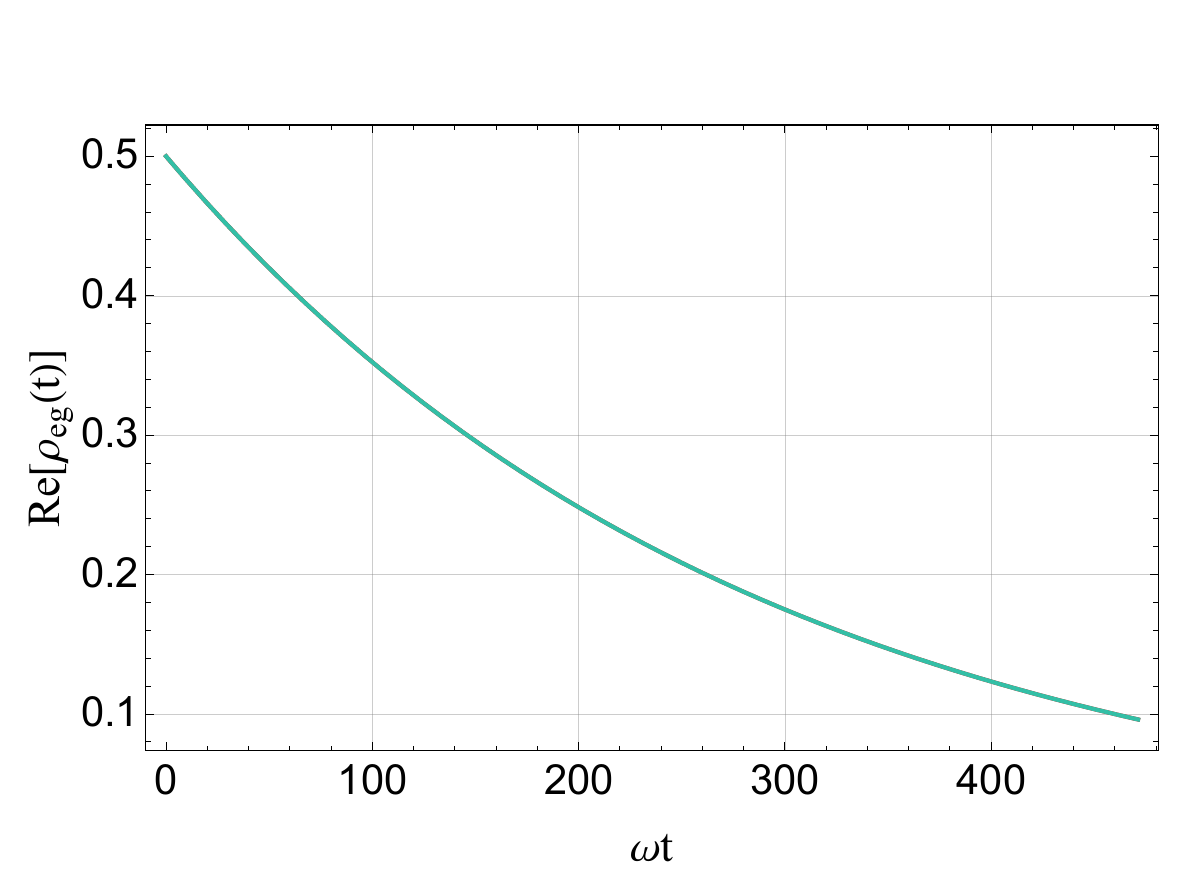}
\caption{Temporal behavior of the populations (panels \textbf{(a)}, \textbf{(c)} and \textbf{(e)}), and coherences (panels \textbf{(b)}, \textbf{(d)} and \textbf{(f)}) of the system's density matrix in different dynamical regimes. Panels \textbf{(a,b)} correspond to the qubit interacting solely with a harmonic oscillator initially prepared in the vacuum state ($p=0$), while panels \textbf{(c,d)} refer to the same configuration with the oscillator prepared in a number state with $p=5$. Panels \textbf{(e,f)} show the case in which the qubit is decoupled from the oscillator ($J=0$) and interacts only with the bath, with coupling strength $\Gamma = 10^{-3}$.
Unless otherwise specified, all simulations are performed with $\omega= 1$, $\omega_D = 0$, and $\lambda = 0$. To focus on each case, we used different time scales by setting $\omega t_f=25\pi$, $\omega t_f=50\pi$ and $\omega t_f=150\pi$ respectively for panels \textbf{(a-b)}, \textbf{(c-d)} and \textbf{(e-f)}.}
\label{fig:specialcase}
\end{figure*}

The first case we are going to analyze is the dynamics of the two-level system interacting only with the harmonic oscillator. 
Let us start by assuming zero excitations in the harmonic oscillator, setting $p=0$, i.e. to consider it in the vacuum state.
In this case, the rates in Eq.~\eqref{eq:rate1}, Eq.~\eqref{eq:rate2} and Eq.~\eqref{eq:rate3} greatly simplify, with $\gamma_1 (t)=2J\tan(Jt)$ and $\gamma_2(t)=\gamma_3(t)=0$.

We recognize in Fig.~\ref{fig:specialcase} {\bf(a)-\bf(b)} the oscillatory dynamics typical of the qubit interacting with a harmonic oscillator.
In detail, the fact that $\rho_{gg}(t) \ge 0.5 $ and $\rho_{ee}(t) \le 0.5 $ indicates that the two-level system tends to lose energy and decays towards the ground state, or, at most, it remains in a balanced superposition of excited and ground state $|\pm\rangle = (|e\rangle \pm|g\rangle )/\sqrt{2}$.

By contrast, it is clear that there is a lack of any beats in the dynamics. This detail further confirms our idea, according to which the beats arise from the interference of oscillations of different frequencies, namely $\Omega_{p+1}=2J \sqrt{p+1}$ and $\Omega_{p}=2J \sqrt{p}$. With $p=0$, the latter clearly vanishes, leaving only the oscillation at frequency $\Omega_{p+1}=2J$ available for our system. It is easy to quantitatively check this assumption from the graphs. For instance, we notice that populations, in Fig.~\ref{fig:specialcase} {\bf(a)}, oscillate with a period $t=2 \pi/ 2J \approx 15 $, in perfect agreement with our prediction. A similar description can be done for coherences, taking into account their different analytical expression.
 
If we combine the observations that can be extrapolated from the plots in Fig.~\ref{fig:specialcase} {\bf(a)} and  Fig.~\ref{fig:specialcase} {\bf(b)}, we can say that the state of the qubit oscillates between the pure state $|g \rangle \langle g|$ and the \textit{non-thermal states} $| \pm \rangle \langle \pm|$, in which the populations are balanced, but the coherences periodically alternate between their maximum and minimum value.
However, these states are still pure states as can be proved by noticing that $\Tr({(|\pm \rangle \langle \pm|)^2})=1$.

We now turn to the general case in which $p \neq 0$. The simulations of this case can be found in Fig.~\ref{fig:specialcase} {\bf(c)-(d)}. {From here, we also see that, both populations and coherences exhibit oscillations with the usual beating.
However, the nodes in the populations pattern appear remarkably narrower than those observed in Fig.~\ref{fig:evolution}. This can be traced back mainly to the absence of the reservoir, whose presence in the general case gives rise to phenomena of dissipation and decoherence. Indeed, the \textit{junction} being broader signals an inversion of populations between the qubit's states $| e \rangle$ and  $| g \rangle$, which is not complete as it is in Fig.~\ref{fig:specialcase} {\bf(c)}. 
The coherences in Fig.~\ref{fig:specialcase} {\bf(d)}, instead, present undamped oscillations throughout the evolution. This behavior further highlights the role of the bath as the sole mechanism responsible for decoherence in the full model.}

Another interesting case concerns the time evolution of the qubit when it is decoupled from the harmonic oscillator.
By imposing the conditions $J=0$, the time-dependent part of the rates Eq.~\eqref{eq:rate1}, Eq.~\eqref{eq:rate2} and Eq.~\eqref{eq:rate3} now becomes identically zero and the equation reduces to
\begin{equation}
\begin{aligned}
	\dot{\rho}_S(t) &=  \Gamma(n+1)\bigg(\sigma_-\rho_S(t)\sigma_+-\frac{1}{2}\big\{ \sigma_+\sigma_-,\rho_S(t)\big\} \bigg) \notag \\
    &+\Gamma n\bigg(\sigma_+\rho_S(t)\sigma_- -\frac{1}{2}\big\{ \sigma_-\sigma_+,\rho_S(t)\big\} \bigg).
	\label{eq:envME_noho}
\end{aligned}    
\end{equation}
We can easily solve this equation, and to have a more precise view of the dynamics under these conditions, we simulate the qubit's behavior for longer time intervals, setting the final time instant $t_{f} = 150 \pi$.
The results shown fully reflect what we would expect for a Markovian dynamics of a qubit interacting with a simple, memoryless environment.
It can be noticed, indeed, that both the populations and the coherences tend to a constant value, which identifies the features of an equilibrium steady state. 
In particular, we observe from Fig.~\ref{fig:specialcase} {\bf(e)} that the populations evolution slows down progressively. We could say that $\rho_{ee}(t)$ and $\rho_{gg}(t)$ reach, in the the limit $t \to \infty$, a \textit{stationary} value.

For what it regards, instead, the real part of coherences, we see in Fig.~\ref{fig:specialcase} {\bf(f)} that they rapidly decay, approaching zero. Such a behavior reflects the characteristics of an asymptotic state where the weight of the off-diagonal elements is strongly downsized. It represents nothing more than the decoherence effect of the bath.
This is not surprising. Indeed, following \cite{Breuer}, it is possible to show that, in absence of external time-dependent fields and for any initial state $\rho_S(0)$, the system tends to an equilibrium, thermal state, in the limit $t \to \infty$. This state is clearly characterized by null coherences and constant value of populations.     

\end{document}